\documentclass[12pt,letterpaper]{article}
\usepackage[a4paper, total={7in, 10in}]{geometry}

\usepackage{graphicx}
\usepackage{helvet}

\usepackage{indentfirst}
\usepackage{authblk}

\usepackage{amsmath} 
\usepackage{amssymb} 
\usepackage{orcidlink}
\usepackage{setspace}
\usepackage[super,sort&compress]{natbib}
\setcitestyle{comma,super}
\usepackage{hyperref}
\usepackage{pdfpages}

\usepackage{subcaption}%

\makeatletter

\newcommand{\Rmnum}[1]{\expandafter\@slowromancap\romannumeral #1@}
\renewcommand\@alph[1]{\number #1}
\makeatother

\makeatletter
\renewcommand{\maketitle}{\bgroup\setlength{\parindent}{0pt}
\begin{flushleft}
  {\LARGE \textbf{\@title}}
  
  \@author
  
\end{flushleft}\egroup}
\makeatother

\title{Decoding RR Lyrae light curves with deep learning for accurate absolute magnitude estimation}

\author[1,2]{ Shunxuan He,}
\author[3,*]{ Huiwen Wu,}
\author[1,2,*]{ Yang Huang,}
\author[3]{ Deyi Zhang,}
\author[3]{ Guirong Xue,}
\author[2,*]{ Jifeng Liu,}
\author[2]{ Xiaodian Chen,}
\author[3]{ Xinyu Qi,}
\author[3]{ Xiaoyu Tang}

\affil[1]{School of Astronomy and Space Science, University of Chinese Academy of Sciences, Beijing 100049, China}

\affil[2]{National Astronomical Observatories, Chinese Academy of Sciences, Beijing 100101, China}

\affil[3]{Research Center for Scientific Data Hub, Zhejiang Laboratory, Hangzhou 310011, China}

\affil[*]{Correspondence: whw@zhejianglab.org (H.W.); huangyang@ucas.ac.cn (Y.H.); jfliu@nao.cas.cn (J.L.)}
\begin{document}

\maketitle

% --- 1. Graphical Abstract (必需项) ---
\section*{GRAPHICAL ABSTRACT}
\begin{figure}[htbp]
    \centering
    \includegraphics[width=1.0\textwidth]{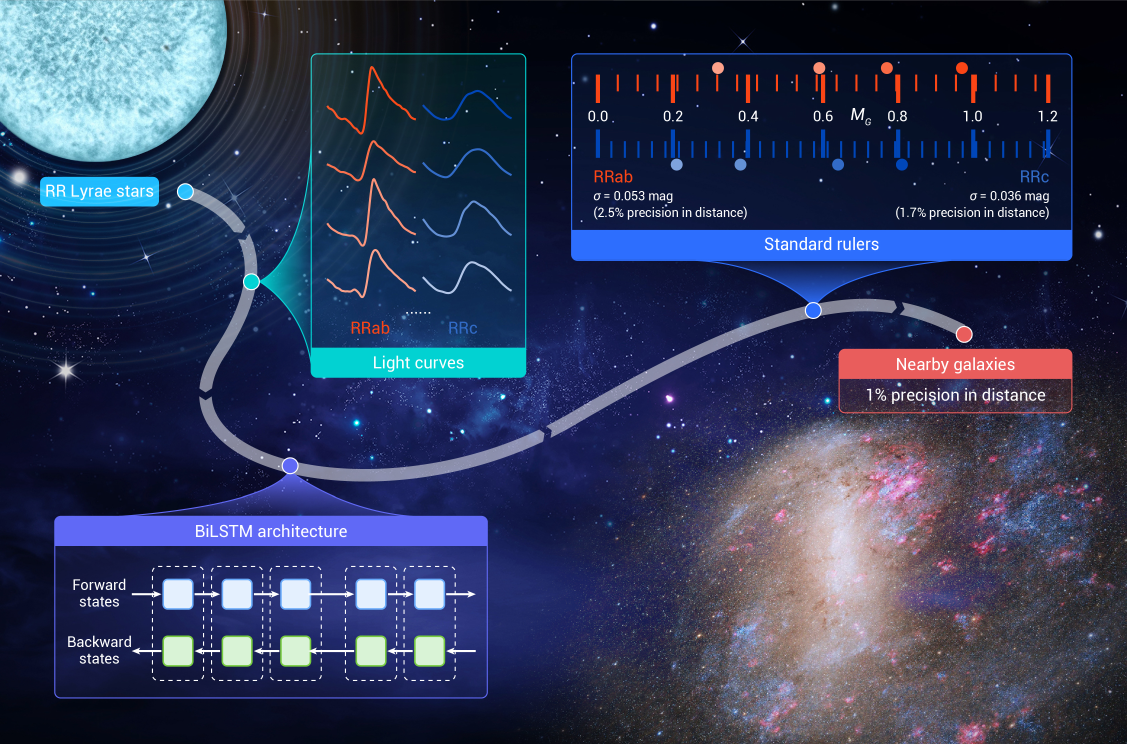} 
\end{figure}

\vspace{0.3cm}

% --- 2. Public Summary---
\section*{PUBLIC SUMMARY}
RR Lyrae stars are pulsating variable stars that serve as standard candles for measuring cosmic distances.

A deep-learning model trained on 80,000 RR Lyrae light curves predicts distances without metallicity.

It achieves 1\% distance precision for Milky Way mapping and independent Hubble-constant constraints.

\newpage

\maketitle

% \section*{ABSTRACT}

RR Lyrae stars are essential standard candles for distance measurements in the Milky Way and nearby galaxies. Traditional estimates rely on the Period--Absolute Magnitude--Metallicity relation but are limited by uncertainties in metallicity determinations. We present a deep learning approach that directly predicts absolute magnitudes from RRab and RRc light curves, eliminating the need for metallicity estimates. Our model achieves validation precisions of 0.053 mag and 0.036 mag (approximately 2.5\% and 1.7\% in distance) for individual RRab and RRc stars, respectively. Tests on globular clusters yield typical distance precisions of 1.0\% for RRab and 1.7\% for RRc stars. For the benchmark systems, combining the RRab- and RRc-based models yields distance moduli of $18.498 \pm 0.001_{\rm stat} \pm 0.018_{\rm sys}$\,mag for the Large Magellanic Cloud and $19.564 \pm 0.003_{\rm stat} \pm 0.019_{\rm sys}$\,mag for the Sculptor dwarf spheroidal galaxy. These measurements are in excellent agreement with previous results, achieve a distance precision of approximately 1\%, and represent a 1.8-fold improvement over traditional RR Lyrae calibration relations. Our approach showcases the ability of AI to directly extract key physical parameters from complex, information-rich light curves, resolve degeneracies, and scale to broader applications.

\section*{KEYWORDS}

% %%%  Include up to 10 keywords, separated by commas. 
% %%%  Keywords entered in EM are not carried over; only 
% %%%  keywords included in the main text will be used 
% %%%  in the final article metadata. Please note that 
% %%%  for some journals, keywords are chosen by the editors.

RR Lyrae stars, Light curves, Recurrent neural networks, Distance measurement

\section*{SUBJECT AREA:}

Astrophysics

\section*{INTRODUCTION}

\begin{figure}[!htbp]
\begin{minipage}{1\textwidth}
\centering
\includegraphics[width=0.98\linewidth]{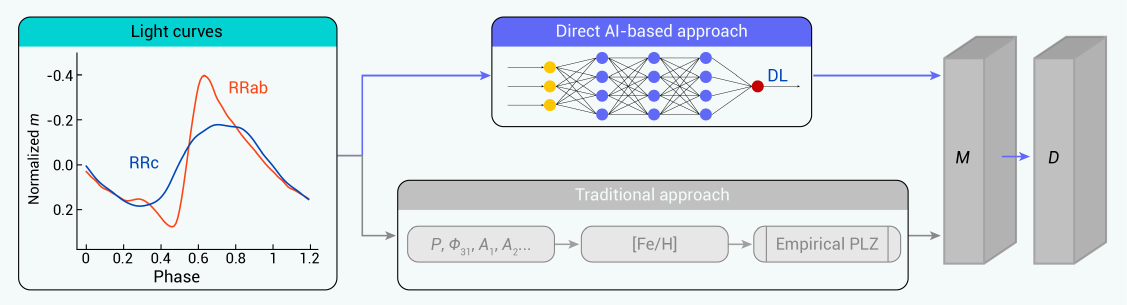}
\end{minipage}
\caption{\textbf{AI-driven estimation of absolute magnitude from light curves.} Our AI-based approach directly infers the absolute magnitude $M_G$ from the light curve, bypassing the traditional two-step procedure: (1) estimating metallicity via photometric calibrations based on Fourier parameters from light curve fitting, and (2) computing $M_G$ using the Period--Luminosity--Metallicity (PLZ) relation.}
\label{fig:AI_for_astro}
\end{figure}

RR Lyrae variable stars, as classical standard candles, play a crucial role in mapping the structure of the Milky Way, measuring distances to nearby galaxies, and serving as potential distance anchors for precise determinations of the Hubble constant. Their distances can be estimated to better than 4\% precision using the calibrated Period--Absolute Magnitude--Metallicity (PMZ) relation.~\cite{Neeley2017,Prudil2024} However, this traditional approach partly depends on accurate metallicity estimates, which are challenging to obtain. Spectroscopic methods are limited by observational depth and spectral resolution, yielding only a limited number of reliable measurements. Photometric approaches rely on empirical relations based on light curve parameterization, introducing additional sources of uncertainty. These dependencies increase methodological complexity and ultimately limit distance precision. A recent study by Chen et al.~\cite{Chen2023} demonstrated that the metallicity of RRd stars (double-mode pulsators) can be robustly inferred from the ratio of their two pulsation periods, offering a promising alternative. However, RRd stars are intrinsically rare, with only a few thousand dentified to date, compared to more than 250{,}000 known RRab (fundamental-mode) and RRc (first-overtone) stars, limiting their utility on a broader scale. In this context, advances in artificial intelligence provide a promising alternative. Here, we propose a deep learning approach that directly predicts absolute magnitudes from RRab and RRc light curves, eliminating the need for prior metallicity estimation, as illustrated in Figure~\ref{fig:AI_for_astro}. By learning a direct mapping from observed variability to intrinsic luminosity, this method simplifies the distance estimation process and has the potential to surpass the precision of traditional metallicity-dependent techniques.

\section*{MATERIALS AND METHODS}

The detailed methods for sample selection, light‑curve preprocessing, model construction and training, member star identification and model testing on globular clusters (GCs) and nearby galaxies, feature ablation analysis, representativeness assessment, and additional analyses of the RRab--RRc model prediction discrepancy are provided in the Supplemental Materials and Methods.

\section*{RESULTS AND DISCUSSION}

We constructed the training dataset by selecting RR Lyrae stars from the Gaia DR3 catalog, utilizing high-quality photometric light curves in the Gaia $G$ band. Leveraging subtype classifications from the Gaia Specific Objects Study (SOS) Cep\&RRL pipeline (ref.~\cite{Clementini23}), we developed separate models for RRab and RRc stars to account for their distinct pulsation characteristics. To ensure the accuracy of the absolute magnitude labels ($M_G$) used for training, we cross-matched the Gaia DR3 sample with the high-precision photometric distance catalog from Ref.~\cite{li2023photometric} (see Supplemental Materials and Methods). Additional quality cuts were applied to the training sample, and stars associated with the Large Magellanic Cloud (LMC), the Sculptor Dwarf Spheroidal Galaxy (Sculptor dSph), and several globular clusters were flagged for independent validation of model performance (see Supplemental Materials and Methods). In total, we obtained 63{,}043 RRab and 16{,}602 RRc stars for model training.

\begin{figure*}[!htbp]
\begin{minipage}{1\textwidth}
\centering
\includegraphics[width=\linewidth]{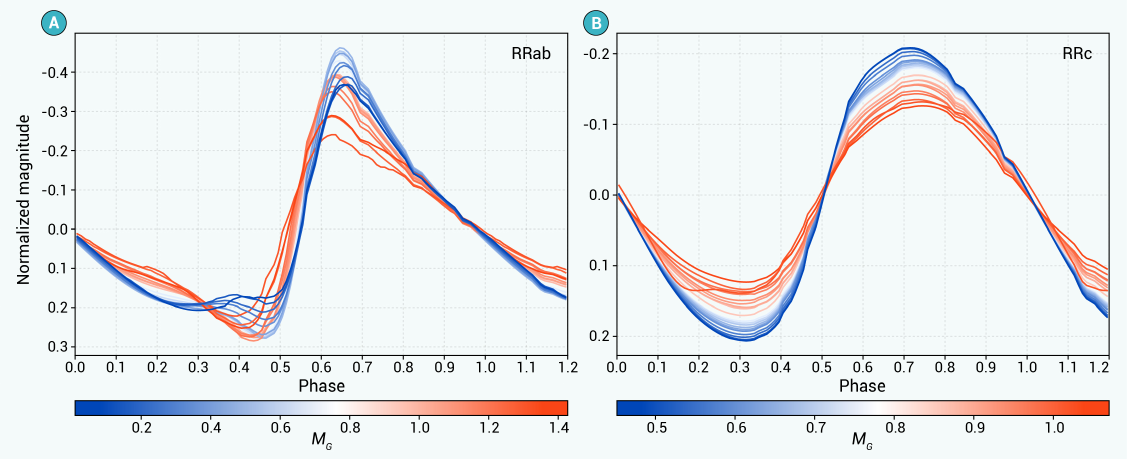}
\end{minipage}
  \caption{\textbf{Typical phase-folded light curves for RR Lyrae stars.} Each curve represents the typical phase-folded light curve corresponding to one of the 20 $M_G$ bins, with phase on the x-axis and mean normalized magnitude (averaged in 0.01-phase intervals) on the y-axis. Curves are color-coded by $M_G$. Panels A and B correspond to RRab and RRc subtypes, respectively.}
  \label{fig:2D_vis}
\end{figure*}

Due to the relatively low cadence and irregular sampling of Gaia photometric data, we transformed the light curves into phase space using Gaia-provided periods and aligned them during pre-processing (see Supplemental Materials and Methods). This process yielded training datasets that preserve the irregularly sampled, variable-length phase-folded light curves derived from the original observations. Figure~\ref{fig:2D_vis} presents the mean phase-folded light curves, grouped into 20 bins based on $M_G$ by using pre-processed light curves. Within each bin, the magnitudes are normalized and averaged in 0.01-phase intervals. Clear morphological trends as a function of $M_G$ emerge for both RRab and RRc stars. These trends confirm that light curves encode rich stellar information beyond what is captured by conventional parameters. Treated as an independent data modality, they allow our model to learn subtle morphological features for accurate magnitude estimation.

Inspired by the systematic dependence of light-curve morphology on absolute magnitude, we develop a sequence-based deep learning framework to predict $M_G$. Given the intrinsically nonlinear relationship between temporal variability patterns and luminosity, models capable of capturing sequential dependencies are essential. Long short-term memory (LSTM) networks~\cite{hochreiter1997long} are well suited for this purpose, as they effectively model long-range correlations in time-series data. Their bidirectional extension (BiLSTM)~\cite{graves2005framewise} further improves representational capacity by incorporating information from both forward and backward temporal directions, and is therefore adopted as our final architecture. To validate the choice of architecture, we systematically benchmark BiLSTM against linear regression,\cite{su2012linear} random forest,\cite{breiman2001random} convolutional neural networks (CNNs),\cite{lecun1989backpropagation,lecun1998gradient} recurrent neural networks (RNNs),\cite{elman1990finding} and Transformer models under identical settings. Both BiLSTM and Transformer outperform all other baseline methods, with BiLSTM achieving performance comparable to Transformer while using substantially fewer parameters. Detailed architecture specifications and benchmarking results are provided in the Supplemental Materials and Methods.

We train the BiLSTM model separately on RRab and RRc datasets. For each dataset, 80\% of the data is used for training and the remaining 20\% for validation. The evolution of the loss function and root mean squared error (RMSE; see Supplemental Materials and Methods) during training demonstrates the model’s convergence and performance on both the training and validation sets. To narrow the gap between training and validation losses and to mitigate overfitting, we employ several regularization techniques, including layer-wise $\ell_2$ regularization, dropout, and early stopping. These strategies enhance the model’s robustness during training. For example, for RRc model, Figures~S2a and~S2b show that both training and validation losses decrease steadily and converge smoothly, indicating minimal overfitting. To further improve generalizability, we perform K-fold cross-validation. As shown in Figures~S2c and~S2d, the training losses and RMSEs across all folds stabilize within a similar range after several epochs, confirming effective learning and consistent performance across different data splits.

We further assess model performance by computing the mean RMSE across five cross-validation folds (see Supplemental Materials and Methods), as summarized in Table~\ref{tab:train_valid}. The close agreement between training and validation RMSEs indicates good generalization capability and minimal overfitting. The RRab model achieves an RMSE of approximately 0.053\,mag, corresponding to a relative distance uncertainty of about 2.5\% for individual stars, whereas the RRc model attains a lower RMSE of approximately 0.036\,mag, equivalent to a distance uncertainty of about 1.7\%. This result suggests that the light-curve–luminosity relation of RRc stars are more readily captured by the model. Predicted and label magnitudes (Figure~S3) follow the one-to-one relation closely, with only small deviations at the bright and faint ends. These deviations are more pronounced for the RRab stars and likely result from a combination of sparse sampling at the edges of the $M_G$ distribution and systematic biases inherited from the photometric metallicity calibration used to derive $M_G$. Consistent with this picture, residuals as a function of $[\mathrm{Fe/H}]$ (Figure~S4) indicate that the RRab model degrades toward both the metal-poor and metal-rich ends, whereas the RRc model remains largely insensitive to metallicity across the sampled range. A more detailed discussion of this effect and its implications for model scalability is provided in the Supplemental Materials and Methods.

\begin{table}[!htbp]
\centering
\caption{Training and validation Root Mean Squared Errors (RMSE) for the BiLSTM model}
\begin{tabular}{l@{\hspace{12pt}}c@{\hspace{12pt}}c}
\\
\hline
Model Type & Train RMSE (mag) & Validation RMSE (mag) \\ 
\hline
RRab & 0.0529 & 0.0534 \\
RRc & 0.0342 & 0.0364 \\ 
\hline
\end{tabular}
\label{tab:train_valid}
\end{table}

\begin{figure}[!htbp]
\begin{minipage}{1\textwidth}
\centering
\includegraphics[width=0.7\linewidth]{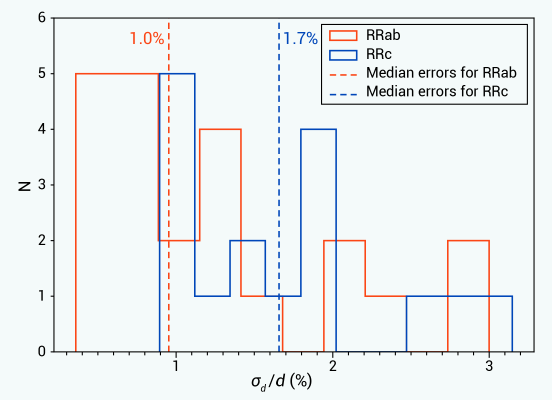}
\end{minipage}
\caption{\textbf{Distribution of relative distance errors for the GCs.} The median relative errors of distance are indicated with red and blue dashed lines for RRab and RRc stars, respectively.}
\label{fig:GCs_precision}
\end{figure}

Building on the trained BiLSTM models, we evaluated their precision using RR Lyrae member stars in GCs. Model performance was quantified in terms of the relative distance precision ($\sigma_d/d$; see Supplemental Materials and Methods for details on distance and uncertainty calculations) for each cluster. The test samples were processed separately by the RRab and RRc models according to their subtype classifications, following the same preprocessing procedures adopted during training. Using distances derived from 22 GCs hosting RRab stars and 16 GCs hosting RRc stars (each with at least five member stars), we achieved typical median distance precisions of 1.0\% for RRab stars and 1.7\% for RRc stars (Figure~\ref{fig:GCs_precision}). The comparatively lower precision of the RRc model is likely attributable to its looser training selection criterion ($M_{G\mathrm{,e}} < 0.2$\,mag, compared to $0.15$\,mag for RRab), which results in larger intrinsic label uncertainties. A detailed analysis of the clusters with larger uncertainties and the factors contributing to these uncertainties is provided in the Supplemental Materials and Methods.

To rigorously assess the accuracy and performance of our models in predicting RR Lyrae distances, we conducted external validations on two benchmark systems: the Large Magellanic Cloud (LMC) and the Sculptor dwarf spheroidal galaxy (dSph). Table~\ref{tab:test_LMCScu} summarizes the distance moduli derived from the RRab and RRc models, with uncertainties that incorporate both statistical and systematic contributions (see Supplemental Materials and Methods for details on distance moduli and uncertainty calculations). We obtain weighted mean distance moduli of $18.498 \pm 0.001_{\mathrm{stat}} \pm 0.018_{\mathrm{sys}}$ for the LMC and $19.564 \pm 0.003_{\mathrm{stat}} \pm 0.019_{\mathrm{sys}}$ for the Sculptor dSph, corresponding to distance precisions of approximately 1\% in both cases. The systematic uncertainties in the weighted mean distance moduli comprise subtype-specific systematic errors from the RRab and RRc models (Table~\ref{tab:test_LMCScu}), together with an additional term accounting for the discrepancy between the two subtypes (see below and Supplemental Materials and Methods). Compared to recent RR Lyrae-based measurements,\cite{MV2016,Garofalo22,Nagarajan2022,Tran2022,li2023photometric} our method yields an overall precision improvement of $\sim$1.8 on average across these two systems. Notably, by leveraging a large sample of RRab and RRc stars within a unified deep learning framework, our approach achieves systematic uncertainties comparable to those of the metallicity-independent RRd-based method of Chen et al.,\cite{Chen2023} while delivering smaller statistical uncertainties due to the substantially larger sample size. This establishes a competitive and broadly scalable alternative for RR Lyrae-based distance determination.

\begin{figure*}[!htbp]
\begin{minipage}{1\textwidth}
\centering
\includegraphics[width=\linewidth]{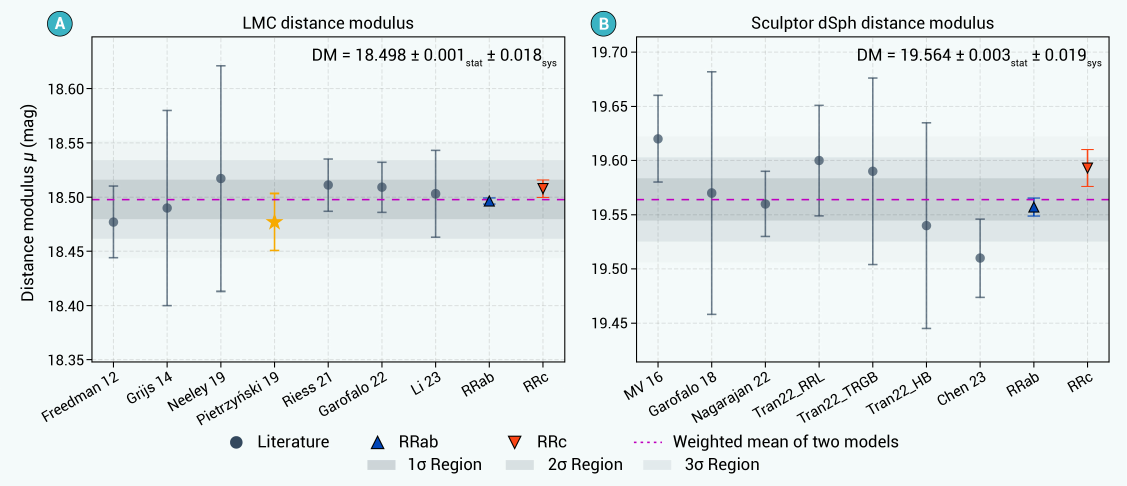}
\end{minipage}
  \caption{\textbf{Comparison of LMC and Sculptor dSph distance moduli between this work and previous studies.} Panels A and B correspond to the LMC and Sculptor dSph, respectively. The horizontal dashed line indicates the weighted average distance modulus derived from our four models, with the shaded gray bands representing the 1$\sigma$, 2$\sigma$, and 3$\sigma$ confidence intervals. Literature values are shown as gray solid circles, and the benchmark LMC measurement from Pietrzyński et al.~\cite{Pietrzynski2019} is highlighted with a gold solid star.}
  \label{fig:lmcscu_compar}
\end{figure*}

To further evaluate accuracy, we compare our results with literature distance measurements based on various independent indicators,\cite{Freedman2012,Grijs2014,Neeley2019,MV2016,Garofalo2018,Pietrzynski2019,Riess2021,Tran2022,Nagarajan2022,Garofalo22,li2023photometric,Chen2023} as shown in Figure~\ref{fig:lmcscu_compar}. For both systems, our predictions, including individual model outputs as well as their weighted averages, are in close agreement with previous measurements and confirm the reliability of our method. In particular, the derived LMC distance modulus is in good agreement with the most precise measurement to date by Pietrzyński et al.,\cite{Pietrzynski2019} differing by only 0.021\,mag. This difference corresponds to approximately 0.7 $\sigma_{\mathrm{tot}}$ (including both statistical and systematic uncertainties) and remains well within the overall error budget. The level of agreement is primarily limited by systematic uncertainties.

\begin{table}[!htbp]
\centering
\caption{Distance Modulus of LMC and Sculptor dSph estimated by RRab/c stars}
\begin{tabular}{l@{\hspace{10pt}}c@{\hspace{10pt}}c@{\hspace{10pt}}c@{\hspace{10pt}}c}
\\
\hline
Model Type & $\mu_\textrm{LMC}$ & N$_\textrm{LMC,\,use}$ & $\mu_\textrm{Sculptor}$ & N$_\textrm{Sculptor,\,use}$ \\ 
\hline
RRab & 18.497$\pm$0.001$_\mathrm{stat}$$\pm$0.003$_\mathrm{sys}$ & 17,485 & 19.557$\pm$0.004$_\mathrm{stat}$$\pm$0.007$_\mathrm{sys}$ & 262\\
RRc & 18.508$\pm$0.002$_\mathrm{stat}$$\pm$0.008$_\mathrm{sys}$ & 6,716 & 19.593$\pm$0.007$_\mathrm{stat}$$\pm$0.016$_\mathrm{sys}$ & 158\\ 
\hline
\end{tabular}
\label{tab:test_LMCScu}
\end{table}

Beyond comparisons with external literature values, examining the internal consistency between the RRab and RRc models provides additional insight into their reliability. As shown in Table~\ref{tab:test_LMCScu}, the distance moduli predicted by the RRab and RRc models differ by roughly 0.01\,mag for the LMC and 0.036\,mag for the Sculptor dSph. These modest systematic differences likely arise from two main factors: (1) target-dependent spatial effects, which are relevant primarily for spatially extended systems such as the LMC, including variations in line-of-sight depth and extinction experienced by RRab and RRc stars due to their non-uniform distribution; and (2) systematic offsets in the training labels, which naturally propagate into the model predictions. The latter is the origin of the additional term included in the systematic uncertainties of the weighted mean distances reported above. In Sculptor dSph, the second factor appears to dominate, providing a clearer view of the models’ intrinsic performance. In the LMC, however, the two effects may counteract one another, leading to a smaller observed offset and an unexpectedly low systematic bias. A more detailed discussion for model discrepancies and error estimation are provided in the Supplemental Materials and Methods.

\begin{figure*}[!htbp]
\begin{minipage}{1\textwidth}
\centering
\includegraphics[width=0.98\linewidth]{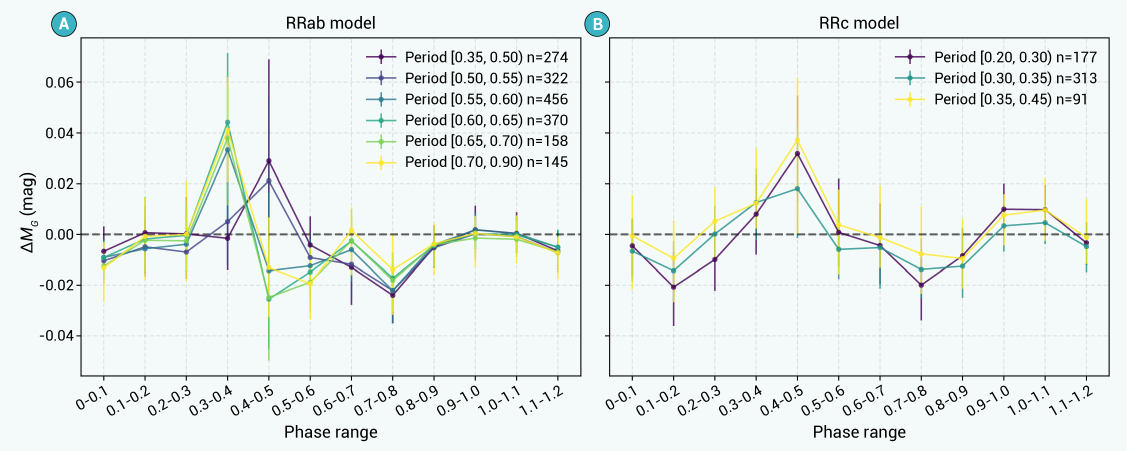}
\end{minipage}
\caption{\textbf{Phase-wise ablation analysis of the RRab and RRc models.} Panel A for RRab and panel B for RRc. Markers indicate the median difference in predicted magnitude ($\Delta M_G$) after ablating each phase interval, with error bars representing the interquartile range (IQR). Colors denote different period bins. Larger absolute values of $\Delta M_G$ reflect greater importance of that phase region.}
\label{fig:phase_import}
\end{figure*}

To further investigate how the model extracts physical information from light curves, we performed a feature ablation analysis. Specifically, data points within sliding phase windows of fixed width were systematically removed from the input, and the resulting changes in predicted magnitudes were used to quantify the importance of each phase region (see Supplemental Materials and Methods for details). In addition to the phase-dependent ablation experiments, we further divide the sample into period bins, motivated by the period-luminosity relation of RR Lyrae stars, to better examine systematic trends. The prediction differences across these period bins are shown in Figure~\ref{fig:phase_import}. Combined with the light-curve morphology presented in Figure~\ref{fig:2D_vis}, these results allow us to interpret the distinct, subtype-dependent patterns. For RRab stars, the bump rise region (phase $\sim$0.3--0.4) is most informative at longer periods, whereas the minimum-light region (phase $\sim$0.4--0.5) becomes more important at shorter periods; the first half of the descending branch (phase $\sim$0.7--0.8) remains consistently influential across all period bins. For RRc stars, the first half of the ascending branch (phase $\sim$0.4--0.5) dominates the prediction, with additional contributions from the left side of the minimum-light region (phase $\sim$0.1--0.2) and the right side of the maximum-light region (phase $\sim$0.7--0.8). These results indicate that the model captures physically meaningful, phase- and period‑dependent features, providing a complementary perspective to traditional Fourier-based approaches for RR Lyrae analysis.\cite{dekany2021,Muraveva2025}

Finally, to place these results in a broader context, we assess the robustness and scalability of the model by examining the representativeness of the test samples and key factors affecting prediction reliability, including parameter distributions, data quality, and modeling assumptions. Detailed analyses are presented in the Supplemental Materials and Methods.

\section*{CONCLUSIONS}

Based on our training and evaluation, we demonstrate the feasibility of directly inferring absolute magnitudes from RR Lyrae light curves using deep learning. Our models achieve distance precisions of 2.5\% and 1.7\% for individual RRab and RRc stars, respectively, approaching the theoretical limits reported by Chen et al.~\cite{Chen2023} In external tests, we reach 1.0\% and 1.7\% distance estimation precision for RRab and RRc stars in globular clusters, respectively. For the LMC and Sculptor dSph, our models deliver $\sim$1\% precision and distance moduli consistent with previous high-precision measurements. These results highlight the potential of applying deep learning directly to observed light curves as a powerful alternative to traditional empirical methods, offering competitive precision and a fundamentally different, data-driven approach to distance estimation.

While our models already perform well, further improvements are possible. Larger training sets and higher-cadence data from \textit{Gaia},\cite{Gaia2023} LSST,\cite{Ivezic2019} and SiTian~\cite{Liu2021,Hemin2025,Huang2025,Han2025} will enhance accuracy and reduce RRab/RRc discrepancies. A promising direction is to train models on multi-band light curves simultaneously, leveraging color and extinction information to better constrain absolute magnitudes and further mitigate systematic uncertainties. Advances in AI combined with detections of more distant RR Lyrae stars will enable precise light-curve distances, crucial for mapping the Milky Way’s 3D structure and studying galaxy evolution. RR Lyrae also provide independent checks on Cepheid and TRGB distance ladders, refining Hubble constant measurements, particularly in the deep time-domain era of LSST. More broadly, our work illustrates how AI can overcome degeneracies inherent in traditional methods by directly learning complex mappings from raw observational data, offering insights for a wide range of astrophysical inference problems.

\section*{RESOURCE AVAILABILITY}

\subsection*{Materials availability}
This study did not generate new unique materials/reagents.

\subsection*{Data and code availability}
Data and code are available from the corresponding author upon reasonable request. The Gaia data used in this work are available from the Gaia archive \href{https://gea.esac.esa.int/archive/}{https://gea.esac.esa.int/archive/}.

\section*{FUNDING AND ACKNOWLEDGMENTS}
This work is supported by the National Natural Science Foundation of China (NSFC grant No. 12422303), the Zhejiang Province Key Research and Development Plan (grant No.~2024SSYS0006), the Fundamental Research Funds for the Central Universities (grant Nos. 118900M122, E5EQ3301X2, and E4EQ3301X2), and the National Key R\&D Programme of China (grant No. 2023YFA1608303). The funders had no role in study design, data collection and analysis, decision to publish, or preparation of the manuscript. We thank Zhirui Li for his valuable assistance in preparing the figures for this manuscript.

\section*{AUTHOR CONTRIBUTIONS}
S.X.H. prepared the data, checked the model accuracy, and drafted the manuscript.
H.W.W. led the model development and training.
Y.H. conceived and designed the project, participated in all analyses, and refined the manuscript.
J.F.L. co-designed the project, participated in all analyses, and refined the manuscript.
D.Y.Z. contributed to the model training.
G.R.X., J.F.L., X.D.C., X.Y.Q., and X.Y.T. provided valuable discussions and revised the manuscript.
All authors contributed to the manuscript and approved the final version.

\section*{DECLARATION OF INTERESTS}
The authors declare no competing interests.

\section*{SUPPLEMENTAL INFORMATION}

It can be found online at xxx.

\renewcommand{\refname}{REFERENCES}
\bibliographystyle{elsarticle-num}
\bibliography{Innovation}

\end{document}

\pagenumbering{gobble}

% ============================================================
% 标题
% ============================================================
\begin{center}
    \textbf{\Large Supplemental Information}\\[2em]
\end{center}
\vspace{2em}

\begin{center}
    \textbf{\Large Table of Contents}
\end{center}

\vspace{1em}

\noindent
\textbf{SUPPLEMENTAL MATERIALS AND METHODS }\\
\textbf{SUPPLEMENTAL FIGURE 1-6 AND FIGURE LEGENDS}\\
\textbf{SUPPLEMENTAL TABLE 1-2}\\
\textbf{SUPPLEMENTAL REFERENCES}\\

\vspace{2em}

\newpage

% ============================================================
% SUPPLEMENTAL MATERIALS AND METHODS
% ============================================================
\noindent\textbf{SUPPLEMENTAL MATERIALS AND METHODS}
\vspace{1em}

\subsubsection*{RR Lyrae training Sample, labels, and selection criteria}\label{sm:trainlabel}

Gaia DR3 has identified approximately 270,000 RR Lyrae star candidates based on multi-band observations since 2013. Among these, `RRab' type stars constitute 64.5\% and `RRc' type stars make up 34.6\% of the total sample, providing a robust dataset for separate training of each subtype.

The training labels, specifically the absolute magnitudes in the $G$ band ($M_G$), are derived from the $M_G-\mathrm{[Fe/H]}$ relation calibrated by Li et al.\cite{li2023photometric} Utilizing the photometric metallicities calculated in their study, distance estimates were provided for 115,410 RRab stars and 20,463 RRc stars. Cross-matching the sample from Li et al.~\cite{li2023photometric} with the Gaia DR3 RR Lyrae sample, we further refined our selection to ensure high-quality and pure sample stars by applying the following criteria: 

\begin{itemize}
\item Each star must have more than 20 observational epochs ($N_\text{ep}$);
\item The phase coverage (Phcov) should be greater than 0.85; 
\item The signal-to-noise ratio (SNR) of the light-curve fitting must exceed 50;
\item The renormalized unit weight error (RUWE) should be less than 1.4;
\item The peak-to-peak amplitude (Totamp) must be within 0.2 to 1.4 for RRab stars and between 0.2 to 0.6 for RRc stars;
\item The absolute magnitude ($M_G$) should range from 0 to 1.5, with corresponding errors smaller than 0.15\,mag for RRab stars and 0.20\,mag for RRc stars.

\end{itemize}

Specifically, the first three criteria ensure that the light curves have high photometric quality and capture detailed morphological features; the fourth and fifth exclude potential contaminants such as eclipsing binaries and misclassified stars; and the final criterion ensures that the training labels are sufficiently precise to constrain the model effectively. For RRc stars, a looser error threshold was adopted to retain a more complete label distribution, since sources with $M_G < 0.45$ (corresponding to [Fe/H] $\sim -2$) typically exhibit uncertainties exceeding 0.15\,mag. The indicators Phcov, SNR, and Totamp, which characterize the light curves, are derived during the pre-processing stage. Here, Phcov quantifies the phase coverage of the raw phase-folded light curve, Totamp denotes the peak-to-peak amplitude of the fitted model, and SNR reflects the quality of the model fit.

Additionally, stars identified as members of globular clusters (GCs) or located in the Large Magellanic Cloud (LMC) or the Sculptor Dwarf Spheroidal Galaxy (Sculptor
dSph) were excluded for extra testing. The complete filtering criteria are detailed in the following section.

After applying the above cuts, we obtained a clean training sample of 63{,}043 RRab and 16{,}602 RRc stars, which were used for model training.

Although the training labels are derived from photometric $M_G-\mathrm{[Fe/H]}$ relations, our models are trained directly on phase-folded light curves rather than on Fourier parameters or the explicit functional form of the $M_G-\mathrm{[Fe/H]}$ relations used to generate the labels. This design ensures that the input representation differs fundamentally from the quantities used in label construction, requiring the model to learn a direct mapping from the light-curve morphology to absolute magnitude. As a result, the risk of methodological circularity is minimized, allowing the model to constrain $M_G$ using only the information encoded in the light curves.

\subsubsection*{The pre-processing of light curves}\label{sm:prepro}

We began by phase-folding the RR Lyrae light curves using periods provided by Gaia. The light curves were then pre-processed using the \textsc{lcfit} package,\cite{istvandekany_2022} which combines Gaussian Process Regression (GPR) with Fourier decomposition to capture variability while mitigating overfitting and ensuring robust performance for sparsely sampled data. Each phase-folded light curve was initially decomposed using a low-order Fourier series:
\begin{equation}
    m(t) = m_{0} + \sum_{i=1}^{k} A_{i} \cos\left(2\pi i t / P + \phi_{i}\right)\text{,}
\end{equation} where $m(t)$ represents the observed magnitude at time $t$, $m_0$ is the mean magnitude, $A_i$ and $\phi_i$ are the amplitude and phase of the $i$-th harmonic, respectively, and $P$ denotes the period. The optimal order $k$ (typically 4--6) was selected by minimizing the mean squared error of the fit. This decomposition yielded Fourier parameters along with their uncertainties. Subsequently, each light curve was normalized by subtracting $m_0$ from $m(t)$, and then phase-shifted and aligned using $\phi_1$ as follows (see also Dékány et al.~\cite{dekany2022photometric}):
\begin{equation}
\label{eqn:phase}
    \mathrm{Phase} = \mathrm{mod}[(t+P\phi_1/(2\pi))/P]\text{.}
\end{equation} 
Following this initial alignment, GPR was applied to generate synthetic light curves sampled at 120 evenly spaced phase points over the interval [0, 1.2], where the range [1.0, 1.2) mirrors [0.0, 0.2) to ensure phase continuity. These synthetic curves were subsequently refitted with a higher-order Fourier series to refine $\phi_1$, which was then used to realign the original observations and produce the final phase-folded light curves used in this work. In parallel, several quality indicators were computed for sample selection. The Phcov was defined as one minus the largest phase gap between adjacent observations. The Totamp was measured from the synthetic light curve. The SNR was computed as:
\begin{equation}
\label{eqn:SNR}
    \mathrm{SNR} = \mathrm{Totamp} \times \sqrt{N_{\mathrm{ep}}} / \sigma_{\mathrm{fit}},
\end{equation}
where $N_\mathrm{ep}$ denotes the number of photometric observations used in the fitting, and $\sigma_\mathrm{fit}$ is the standard deviation of the residuals between the model and the data. Further details can be found in Table~1 of He et al.~\cite{He2025}

Furthermore, we emphasize that Fourier parameters are not used as input features in model training, despite their role in preprocessing. Their sole purpose is to determine the phase alignment through $\phi_1$, enabling consistent comparison of magnitudes across light curves. This alignment, performed in Fourier space, is conceptually similar to using epochs of maximum or minimum light for alignment in the time domain; although the implementation differs, the two approaches are essentially equivalent. Additional tests show that models trained with either alignment approach yield nearly identical RMSE, indicating that the choice of alignment has negligible impact on performance. To ensure internal consistency, we do not adopt Gaia’s maximum-light epochs, as they are derived from $\phi_{\mathrm{max}}$ obtained via Gaia’s Fourier fitting and may introduce systematic offsets. For future applications with high-cadence and high-precision data, such as from TESS or Kepler, maximum-light alignment may provide a more stable phase reference and could further improve model performance.

%% models 
\subsubsection*{Detailed model construction}\label{sec:model_cons}

In this section, we briefly describe the training model used to predict the absolute magnitude $M_G$.  
The input to the model is light-curve data with shape $(N, 2)$, where $N$ is the number of data points in the light curve,  
and the two columns correspond to the phase and magnitude, respectively.  
We denote the two-dimensional sequential light-curve input as:
\[
\mathbf{X} = \begin{bmatrix} x_1, x_2, \cdots, x_N \end{bmatrix}
= \begin{bmatrix} 
P \cdot p_1 & P \cdot p_2 & \cdots & P \cdot p_N \\
m_1 & m_2 & \cdots & m_N \\
\end{bmatrix},
\]
where $p_i$ is the phase value computed as in Eq.~\eqref{eqn:phase}, and $m_i$ is the normalized magnitude, and $P$ is the pulsation period provided by Gaia. The phase is explicitly scaled by the period to incorporate period information into the input representation. We emphasize that the target label ($M_G$) is derived solely from the $M_G-\mathrm{[Fe/H]}$ relation. Therefore, including the period as an input feature does not materially increase the risk of circularity.

Inspired by the method of Dékány and Grebel,\cite{dekany2022photometric} we adopt a bidirectional long short-term memory (BiLSTM) network~\cite{schuster1997bidirectional} to process the input sequential light curve data $\mathbf{X}$.  
The BiLSTM extends the traditional LSTM~\cite{hochreiter1997long} by processing the sequence in both forward and backward directions, enhancing its ability to capture long-range dependencies.

First, we process the input light curve data of shape $(N,2)$ by normalizing the magnitude values, followed by padding (when the input length is shorter than the predefined length) and truncation (when the input length exceeds the predefined length). Then, the standardized light curve feature is fed into the BiLSTM block, which comprises a forward LSTM layer (left-to-right) and a backward LSTM layer (right-to-left). In the forward LSTM, hidden states $h_t^{(f)}$ are computed at each time step $t$, capturing information from the preceding sequence. The forward LSTM includes four gates: the forget gate $f_t^{(f)}$, input gate $i_t^{(f)}$, cell gate $c_t^{(f)}$, and output gate $o_t^{(f)}$. The hidden state $h_t^{(f)}$ is updated based on these gates as follows:
\[
h_t^{(f)} = o_t^{(f)} \circ c_t^{(f)} \circ i_t^{(f)} \circ f_t^{(f)} (
\begin{bmatrix}
x_1, x_2, \cdots, x_{t-1}, x_t 
\end{bmatrix}
),
\]
where $\mathbf{X}$ is the input sequential data of two dimensions.  
In addition to the forward LSTM, a backward LSTM processes the input sequence in the reverse direction, allowing the model to capture future context.  
Similar to the forward LSTM, the backward LSTM comprises the forget gate $f_t^{(b)}$, input gate $i_t^{(b)}$, cell gate $c_t^{(b)}$, and output gate $o_t^{(b)}$.  
The hidden state $h_t^{(b)}$ is computed as:
\[
h_t^{(b)} =
o_t^{(b)} \circ c_t^{(b)} \circ i_t^{(b)} \circ f_t^{(b)} \biggl(
\begin{bmatrix} x_N, x_{N-1}, \dots, x_t \end{bmatrix} \biggr),
\]
where $h_t^{(f)}$ denotes the hidden state from the forward LSTM block.

After we obtained the hidden states from both forward and backward LSTMs, we further concatenate the two hidden states into one tensor. 
\[
h_t = [h_t^{(f)}, h_t^{(b)}], 
\]
where $h_t^{(f)}$ is the hidden state of the forward LSTM block and $h_t^{(b)}$ is the hidden state of the backward LSTM. 
In order to alleviate overfitting issues, we further add a dropout layer after the BiLSTM block. 

In addition, a second BiLSTM layer is applied, followed by another dropout layer. 
The output from this second BiLSTM layer is then passed to a final regression layer, implemented as a dense linear layer.  

\begin{figure}[!htbp]
\begin{minipage}{0.99\textwidth}
\centering
\includegraphics[width=0.45\linewidth]{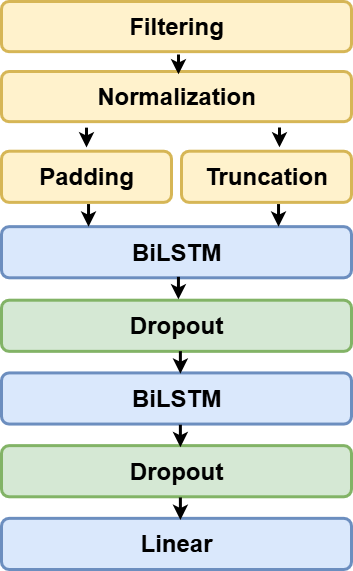}
\end{minipage}
\caption{\textbf{Outline of model structure.} The yellow blocks represent the feature processing pipeline, which includes filtering, normalization, and length adjustment (padding or truncation) to prepare time-series data for the model. The blue blocks correspond to trainable neural network layers, consisting of two bidirectional LSTM layers followed by a linear output layer, which capture temporal patterns from the processed data. The green blocks denote non-trainable dropout layers, used during training to prevent overfitting by randomly deactivating neurons.}
\label{fig:model_struc}
\end{figure}

The overall model architecture is illustrated in Figure~\ref{fig:model_struc}. 
The total number of parameters in the model is 134{,}273. 
The model is implemented using the \textsc{Pytorch} package~\cite{imambi2021pytorch}, 
which provides an efficient and flexible framework for building and training deep learning models.
All models are trained on a system equipped with 2 Nvidia A40 GPUs.

\subsubsection*{BiLSTM training}\label{sm:lstm_train}

Each dataset is split into training and validation sets at a ratio of 8:2, where 80\% of the data is used for training and 20\% for validation. This split is controlled by a fixed random seed of $42$ to ensure reproducibility. Training is conducted using the AdamW optimizer~\cite{loshchilov2017decoupled} with a batch size of 128, and the model is trained for a maximum of 300 epochs. The optimal hyperparameters are listed in 
Table~\ref{tab:hyper}.
% table~S1.
Dim 1 corresponds to the output dimension of the first LSTM layer, while Dim 2 represents the output dimension of the second LSTM layer.
Hyperparameters are selected through a grid search, with Dim 2 varying across $\{32, 64, 128\}$, learning rates of $\{0.0005, 0.001\}$, and weight decay values in $\{0, 10^{-5}\}$. 
Each training procedure is repeated five times, and the final performance metric is reported as the mean across all runs. The raw phase-folded light curve sequences of unequal lengths are standardized by padding shorter sequences with $-1$ or truncating longer sequences to a fixed length of 80 data points for RRab and 70 data points for RRc during model training.
To evaluate the evolution of the training process and assess model convergence, we monitor both the loss function and the root mean squared error (RMSE) throughout training. These metrics are plotted against the number of epochs to illustrate how the model performance improves over time. RMSE is used as the primary accuracy metric for predicting the absolute magnitude $M_G$, measuring the magnitude of prediction errors through the following standard definition:
\begin{equation}
\label{eqn:root_mse}
\sigma = \sqrt{\frac{1}{N} \sum_{i=1}^N (y_i - \hat{y}_i)^2},
\end{equation}
where $N$ is the total number of stars, and $y_i$ and $\hat{y}_i$ denote the true and predicted absolute magnitudes of the $i$-th star, respectively. A smaller RMSE indicates more accurate predictions, as it reflects less dispersion of the residuals. We report RMSE on both the training and validation sets to evaluate model performance.

We further explored several alternative training strategies to assess the robustness of our data representation and model design.
First, we trained the model using synthetic phase-folded light curves generated via GPR and uniform resampling, instead of the raw observations. Under this setting, the RRab model achieved comparable performance, while the RRc model showed only a marginal improvement. We also experimented with combining raw and synthetic light curves as joint inputs for training; however, this approach did not lead to any noticeable performance improvement compared to using either representation alone. Based on these tests, we adopt raw light curves as the primary input, as they retain observational noise and subtle morphological features that are both informative and representative of real data.
Second, we trained a unified model by combining RRab and RRc samples, rather than using subtype-specific models. The resulting performance is comparable to that of the less accurate RRab model, suggesting that the combined model is primarily constrained by the more challenging subtype. This result indicates that, given the current dataset and methodology, merging the two populations does not provide a meaningful improvement in performance.

\subsubsection*{Member stars in GCs/LMC/Sculptor dSph}\label{sm:test_mem}
The member stars of these systems were all selected from the Gaia RR Lyrae catalog. For GCs, we focused on the 157 clusters identified by Harris et al.~\cite{harris2010new} and employed proper motion data (including uncertainties) from Vasiliev \& Baumgardt~\cite{vasiliev2021gaia} to identify GC members. Specifically, the selection criteria were as follows:

\begin{itemize}
    \item The angular separation from the center of the GC is required to be within $15\,r_{\rm h}$, where $r_{\rm h}$ is the half-light radius of the cluster.
    \item The proper motions must satisfy: 
    \[
    |\mu_{\alpha} - \mu_{\alpha, {\rm GC}}| \leq 8\,\sigma_{\mu_{\alpha, {\rm GC}}} \quad \text{and} \quad |\mu_{\delta} - \mu_{\delta, {\rm GC}}| \leq 8\,\sigma_{\mu_{\delta, {\rm GC}}},
    \]
where $\mu_{\alpha}$ and $\mu_{\delta}$ are the proper motions of the star, and $\mu_{\alpha, {\rm GC}}$ and $\mu_{\delta, {\rm GC}}$ are the referenced proper motions of the GC.
\end{itemize}

For these GC member stars, we further refined the test sample by applying additional quality cuts that similar to the selection criteria for the training sample: (1) $N_\mathrm{ep}>20$; (2) $\mathrm{Phcov}>0.85$; (3) $\mathrm{SNR}>30$; (4) $\mathrm{RUWE}<1.4$; (5) $0<M_G<1.5$. To ensure reliable validation for each GC, we required a minimum of five RRab or RRc stars per cluster. After applying these filters, we selected 390 RRab stars from 22 GCs and 152 RRc stars from 16 GCs to test the model performance.

For the LMC, member stars were selected based on the positional and proper motion properties. The center of the LMC is defined as $(\alpha_{\mathrm{C, LMC}}, \delta_{\mathrm{C, LMC}}) = (81.28^{\circ}, -69.78^{\circ})$ as given by van der Marel.\cite{van2001magellanic}

The selection criteria were as follows:

\begin{itemize}
    \item The angular distance from the center of the LMC must be less than $10^{\circ}$.
    \item The proper motions should satisfy:
    \[
    \left[0.128\,(\mu_{\alpha} - 1.80) - 0.681\,(\mu_{\delta} - 0.38)\right)^2 + \left(1.13\,(\mu_{\alpha} - 1.80) + 0.322\,(\mu_{\delta} - 0.38)\right]^2 < 1,
    \]
    where $\mu_{\alpha}$ and $\mu_{\delta}$ are the proper motions in right ascension and declination directions, respectively, in units of mas yr$^{-1}$. This empirical criterion defines a rotated ellipse in proper motion space, similar to the method proposed by Nidever et al.\cite{Nidever2020}
\end{itemize}

Sculptor dSph member stars were selected based on Martínez-Vázquez et al.,\cite{MV2016} who identified them using updated Welch-Stetson variability index,\cite{Welch1993,Stetson1996} light curve properties, and positions in the color--magnitude diagram. We then cross-matched these sources with Gaia DR3 to obtain their photometric data.

For both LMC and Sculptor dSph member stars, we further applied similar quality cuts:  
(1) $N_{\mathrm{ep}} > 20$; (2) $\mathrm{Phcov}> 0.85$; and (3) $\mathrm{RUWE}< 1.4$. In total, we obtained 18,597 RRab and 7,195 RRc type LMC member stars, along with 262 RRab and 158 RRc type Sculptor dSph member stars.

\subsubsection*{Model comparison and selection}\label{sm:model_compar}

To empirically validate the choice of the BiLSTM architecture, we conducted a systematic comparison against several baseline models under identical data, preprocessing, and evaluation protocols. All models were trained and evaluated for RRab and RRc subtypes using the same training/validation split (80/20).

The compared models include:
\begin{itemize}
    \item \textbf{Linear Regression}: A standard linear model fitted to the phase and magnitude vectors, serving as a non-sequential baseline.
    \item \textbf{Random Forest}: An ensemble of decision trees trained on the same flattened input, with the number of trees optimized via grid search.
    \item \textbf{CNN}: A one-dimensional convolutional network with two convolutional layers followed by global average pooling and a dense output layer, designed to capture local morphological features.
    \item \textbf{RNN}: A single-layer recurrent neural network with tanh activation, included to assess the necessity of gated architectures.
    \item \textbf{Transformer}: A sequence-to-value encoder-only model with multi-head self-attention, positional encoding, and a regression head, implemented with a comparable parameter budget to the BiLSTM.
    \item \textbf{BiLSTM}: The proposed two-layer bidirectional LSTM as described in the main text.
\end{itemize}

The validation root mean squared error (RMSE) for each model is summarized in Table~\ref{tab:model_comparison}. BiLSTM significantly outperforms linear regression and random forest by a factor of approximately 2--3 in validation RMSE, while also achieving moderate improvements (20\%--40\%) over CNN and RNN. BiLSTM and Transformer achieve comparable accuracy, with differences well within 0.002\,mag. Notably, BiLSTM reaches this level of accuracy with a substantially smaller parameter budget (approximately 134,000 parameters) compared to the Transformer (over 200,000 parameters), as well as lower computational cost, justifying its use as the primary model in this work.

\subsubsection*{Testing in GCs, LMC, and Sculptor dSph}\label{sm:test_results}

We evaluated the trained BiLSTM models for RRab and RRc stars using independent test samples from GCs, LMC, and Sculptor dSph. For each validation, the input consisted of light curve sequences pre‑processed in the same manner as the training data, and the output was the predicted absolute magnitudes.

To propagate uncertainties from the training labels, we generated 50 perturbed realizations by adding Gaussian noise consistent with the label uncertainties reported by Li et al.\cite{li2023photometric} For each realization, we retrained the BiLSTM models for both RRab and RRc subtypes, yielding 50 independent model instances per subtype. These instances were used exclusively for uncertainty estimation, while the baseline models trained on the original labels were adopted to derive the final distance estimates.

For GC member stars, predicted absolute magnitudes were converted into distances after applying extinction corrections from Harris et al.\cite{harris2010new} Cluster distances were then determined as the mean distance $d$ of member stars based on the baseline model. To estimate statistical uncertainties, we evaluated each of the 50 model instances separately and computed the uncertainty for each instance using $\sigma/\sqrt{N}$ for clusters with fewer than 10 members and bootstrapping for those with 10 or more members, where $\sigma$ is the standard deviation of the predicted distances and $N$ is the number of members. The final statistical uncertainty for each cluster was taken as the average of these 50 estimates. The systematic uncertainty was defined as the standard deviation of the 50 mean cluster distances derived from the different model instances.
The typical precision of each model is quantified as the median total relative uncertainty, $\sqrt{\sigma_{\mathrm{stat}}^2 + \sigma_{\mathrm{sys}}^2}/d$, across all clusters. Clusters with relatively large distance uncertainties, approaching $\sim$3\%, are primarily driven by increased statistical errors, which arise from small sample sizes and, in some cases, lower-quality light curves with reduced SNR, incomplete phase coverage, and larger fitting residuals.

For the LMC and Sculptor dSph samples, we followed an analogous procedure. Absolute magnitudes predicted by the baseline models were converted into distance moduli. For the LMC, extinction values were derived using the extinction map and the color-excess relation $E(B-V) = E(V-I)/1.237$ from Skowron et al.,\cite{Skowron2021} and converted to $A_G = R_G \times E(B-V)$ with $R_G = 2.516$ from Huang et al.\cite{Huang2021} The distance modulus for each star was computed accordingly, and the mean distance modulus for each system was obtained by fitting a Gaussian to the distribution. Statistical uncertainties were estimated in the same way as for the GCs, by averaging the bootstrapped uncertainties derived from each of the 50 model instances, while systematic uncertainties were defined as the standard deviation of the 50 mean distance moduli. The same procedure was applied to the Sculptor dSph sample, adopting reddening values from Schlegel et al.~\cite{SFD98} and applying $3\sigma$ outlier rejection.

To obtain the final combined distance moduli for the LMC and Sculptor dSph, we performed inverse-variance weighted averages of the RRab and RRc baseline results, with uncertainties propagated accordingly. To account for potential systematic offsets between the RRab and RRc models, likely inherited from differences in their training labels,\cite{li2023photometric} we included an additional subtype-dependent systematic term. For the Sculptor dSph, where spatial effects are negligible, this term was estimated as half of the difference between the RRab and RRc distance moduli, approximately 0.018\,mag. For the LMC, the observed offset between the RRab and RRc models may reflect a combination of subtype-dependent systematics and additional effects related to the system’s geometry (e.g., line-of-sight depth) and spatially varying extinction. As these contributions cannot be reliably disentangled with the current data, we do not attempt an explicit decomposition. Instead, we adopt the Sculptor-based value (0.018\,mag) as a proxy for the subtype-dependent systematic uncertainty, providing a conservative estimate of this contribution. This term is then incorporated into the total uncertainty budget of the weighted mean distances. The final results for RRab and RRc model and their uncertainties are summarized in Table~2 and illustrated in Figure~4. We note that any potential global zero-point systematic in the adopted absolute magnitude calibration is not included in the present uncertainty budget and would introduce a coherent shift to all distance estimates.

\subsubsection*{Feature ablation analysis}\label{sm:feature_importance}

To interpret the model predictions and identify which phases of the light curve contribute most to the absolute magnitude estimation, we performed a feature ablation analysis. This analysis was conducted on a high-quality subset of the validation set, selected based on two criteria: (1) the absolute residual between the predicted and label magnitudes is less than 0.01\,mag, ensuring reliable model performance on these samples; and (2) the phase coverage is at least 0.9, guaranteeing adequate sampling across the folded light curve.

For each selected light curve, we systematically removed data points within a sliding phase window of width 0.1 and re-predicted the absolute magnitude using the trained model. The importance of a given phase interval is quantified as the change in prediction after ablation, defined as
\[
\Delta M_G = M_{G,\mathrm{original}} - M_{G,\mathrm{ablated}},
\]
where a larger value of $|\Delta M_G|$ indicates a greater contribution of the ablated phase region to the model prediction.

Because both the period--luminosity relation and the model input encode period information (via the scaled phase $P \cdot p_i$), the importance of each phase region may vary with pulsation period. To account for this, we divided the sample into several period bins. For each phase interval within each bin, we computed the median and interquartile range (IQR) of $\Delta M_G$. The results are presented in Figure~5, where the left and right panels correspond to the RRab and RRc models, respectively. The analysis reveals clear phase- and period-dependent importance patterns, which are discussed in the main text.

\subsubsection*{Assessment of test sample representativeness and model scalability}\label{sm:representativeness}

To evaluate the reliability and scalability of the trained models, we performed a series of diagnostic analyses on the test samples, focusing on parameter distributions, data quality, and underlying assumptions. Figure~\ref{fig:para_distri} shows the distributions of metallicity ([Fe/H], from Li et al.~\cite{li2023photometric}), absolute magnitude ($M_G$), and distance for the training, validation, and test sets, which form the basis for the following assessment.

We first examine the metallicity distribution, which directly affects the absolute magnitude scale. Relative to the training and validation sets, the test sample is moderately shifted toward lower metallicity due to the dominance of extragalactic populations. As indicated by the residual--metallicity relation (Figure~\ref{fig:residual_feh}), the RRab model performs robustly within an intermediate range ($-2.5 \lesssim \mathrm{[Fe/H]} \lesssim 0$), but exhibits systematic deviations toward both metal-poor and metal-rich ends. These deviations are consistent with the behavior seen at the edges of the $M_G$ distribution and likely arise from a combination of reduced training coverage in these regimes and systematic biases propagated from the photometric metallicity calibration used to derive the training labels. We note that this intermediate metallicity range, corresponding to the stable performance regime of the RRab model, already encompasses at least 97\% of the RRab stars in our sample. Because our sample closely follows the intrinsic RRab metallicity distribution, this indicates broad scalability of the model to the majority of the RRab population. 
In contrast, the RRc model remains stable across its full sampled metallicity range ($-3 \lesssim \mathrm{[Fe/H]} \lesssim -0.5$). Since most test sources lie within the well-performing regime, this shift does not significantly affect the representativeness of the test sample.

We then assess the distance distribution. Aside from a small number of nearby globular cluster members, the sample is dominated by more distant systems, primarily the LMC at distances of $\sim$50\,kpc. Despite this difference, high-quality light curves in the test set yield prediction uncertainties comparable to those of the validation set. This is consistent with Figure~\ref{fig:SNR_M_G_e}, where the prediction uncertainty ($e_{M_{G}}$) for high-quality LMC sources remains similar to that of the validation sample, indicating that model performance does not significantly degrade with distance when data quality is sufficient.

To quantify the impact of data quality, we analyze the relation between signal-to-noise ratio (SNR) and prediction uncertainty. The prediction uncertainty systematically grows toward lower-quality data, particularly for SNR $\lesssim$60, as shown in Figure~\ref{fig:SNR_M_G_e} panel (b) for the LMC sample. This trend reflects the strict quality cuts applied to the training data, which introduce a selection bias and limit the model’s ability to robustly separate signal from noise in lower-quality observations. As a result, stars in high-extinction regions, crowded fields, or near the photometric detection limit are more prone to degraded predictions. In this context, the effective distance reach of the model is determined primarily by data quality rather than by distance itself. For stars in regions of low extinction and weak crowding, the model could in principle be applied out to distances of approximately 100\,kpc, assuming a Gaia detection limit of $G\sim20.5$\,mag and a typical absolute magnitude of $M_G\sim0.5$\,mag. In practice, however, this theoretical limit is seldom reached because light-curve quality degrades substantially at low signal-to-noise ratio. Achieving reliable inference in this regime therefore remains a major challenge for current data-driven approaches.

We also assess the impact of period uncertainties. Gaia periods are derived using Lomb–Scargle periodograms~\cite{Lomb1976,Scargle1982} and refined through non-linear fitting.\cite{levenberg1944,marquardt1963} Based on the individual period uncertainties reported by the Gaia SOS Cep\&RRL pipeline,\cite{Clementini23} we find that they are below $10^{-4}$ days in nearly all cases, with median values of order $\sim 10^{-6}$ days for both RRab and RRc stars. This level of precision ensures a negligible impact on phase folding and period scaling. In addition, Gaia’s long temporal baseline and quasi-random time sampling reduce aliasing relative to typical ground-based surveys, whose sampling is often affected by periodic gaps such as day–night cycles and seasonal visibility. Period uncertainties can therefore be safely neglected in both training and prediction.

The current model does not explicitly distinguish Blazhko variables, and their inclusion may introduce additional scatter due to amplitude and phase modulation. Separating Blazhko and non-Blazhko stars may further improve performance in future work.

Finally, the model is trained exclusively on \textit{Gaia} $G$-band photometry. Its scalability to other surveys with different photometric systems, cadence patterns, and noise characteristics is therefore not guaranteed. Extending the model to other datasets would require retraining with survey-specific light curves following a similar framework.

\subsubsection*{Possible origins of the RRab--RRc model prediction discrepancy}\label{sm:ana_modeldiff}

The observed discrepancy between the RRab and RRc model predictions may arise from two different sources: target-dependent effects associated with the validation targets and systematic offsets inherited from the training labels. We discuss them separately below.

The first possible origin is associated with the validation targets themselves. Spatial effects are expected only for spatially extended systems such as the Magellanic Clouds. Owing to the inclined disk structure of the LMC, stars along different lines of sight may exhibit systematic distance variations. Consequently, the non-uniform and distinct spatial distributions of RRab and RRc stars can introduce systematic offsets between their predicted distance moduli. This effect is expected to be even stronger in the SMC, where the smaller and less uniformly distributed RR Lyrae sample introduces more severe sampling biases. This limitation is also the main reason why the SMC was not included in our final validation. In addition, spatially varying extinction introduces further uncertainties. In regions of high extinction, particularly those containing higher concentrations of RR Lyrae stars, errors in the extinction correction may introduce non-negligible biases in the inferred distances.

The second possible origin is the systematic offset inherited from the training labels. In the study by Li et al.,\cite{li2023photometric} distance estimates derived from the Gaia $G$-band $M_G-\mathrm{[Fe/H]}$ relation exhibit a relative offset of approximately 0.8\% between RRab and RRc stars, as validated using globular cluster members. This corresponds to an absolute-magnitude difference of about 0.02\,mag and is therefore inherited by the independently trained RRab and RRc models.

The relative importance of these two effects depends on the validation target. For the compact and more distant Sculptor dSph, spatial effects are expected to be negligible, making the systematic label offset the dominant source of the observed discrepancy between the RRab and RRc model predictions. In contrast, both effects can influence the LMC results. If spatial effects cause the RRab model to predict slightly larger distances and/or the RRc model slightly shorter ones, they may partially compensate for the systematic label offset. As a result, the apparent discrepancy between the two models may be artificially reduced, leading to an underestimated systematic uncertainty. A more reliable assessment of these effects will require larger and more spatially uniform RR Lyrae samples from future time-domain surveys.

\newpage

% ============================================================
% SUPPLEMENTAL FIGURES
% ============================================================
\noindent\textbf{SUPPLEMENTAL FIGURES AND FIGURE LEGENDS}
\vspace{2em}

% done with update --0722
\begin{figure}
  \begin{minipage}{1\textwidth}
    \centering
    \begin{subfigure}{0.475\textwidth}
        \includegraphics[width=\textwidth]{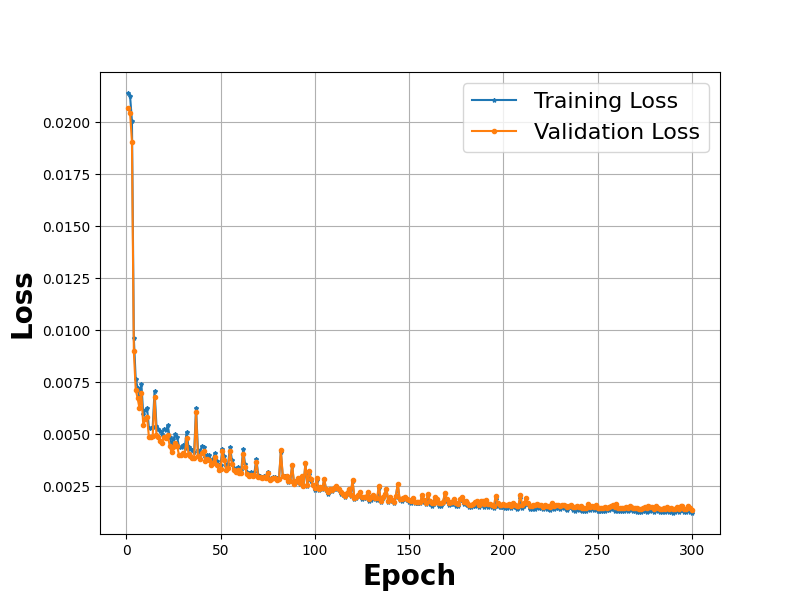}
        \caption{RRc Loss}
        \label{fig:rrc_loss}
    \end{subfigure}
    \begin{subfigure}{0.475\textwidth}
        \includegraphics[width=\textwidth]{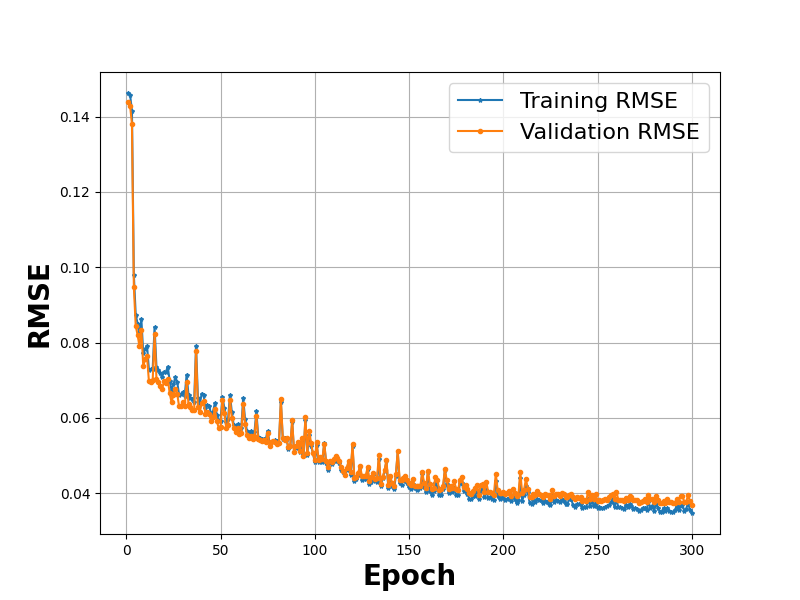}
        \caption{RRc RMSE}
        \label{fig:rrc_rmse}
    \end{subfigure}
    \begin{subfigure}{0.475\textwidth}
        \includegraphics[width=\textwidth]{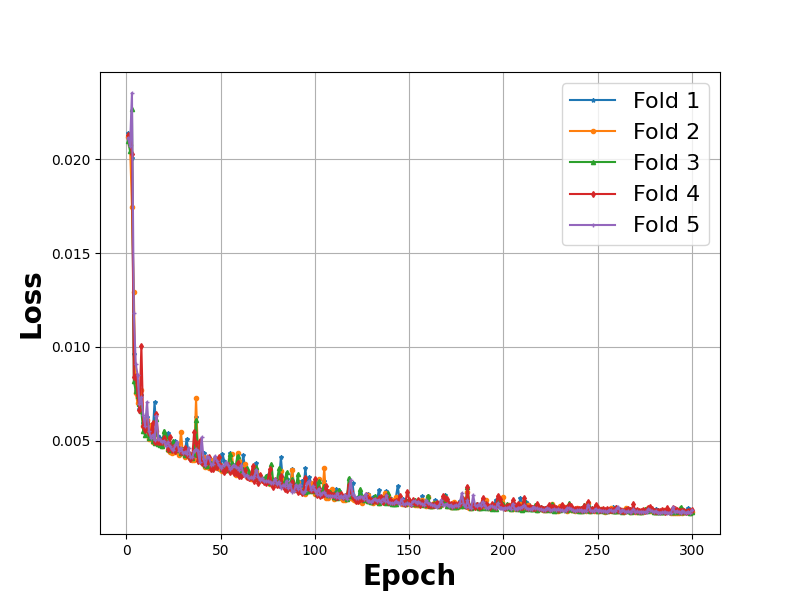}
        \caption{RRc Train Loss K-folds}
        \label{fig:rrc_loss_multi}
    \end{subfigure}
    \centering
    \begin{subfigure}{0.475\textwidth}
        \includegraphics[width=\textwidth]{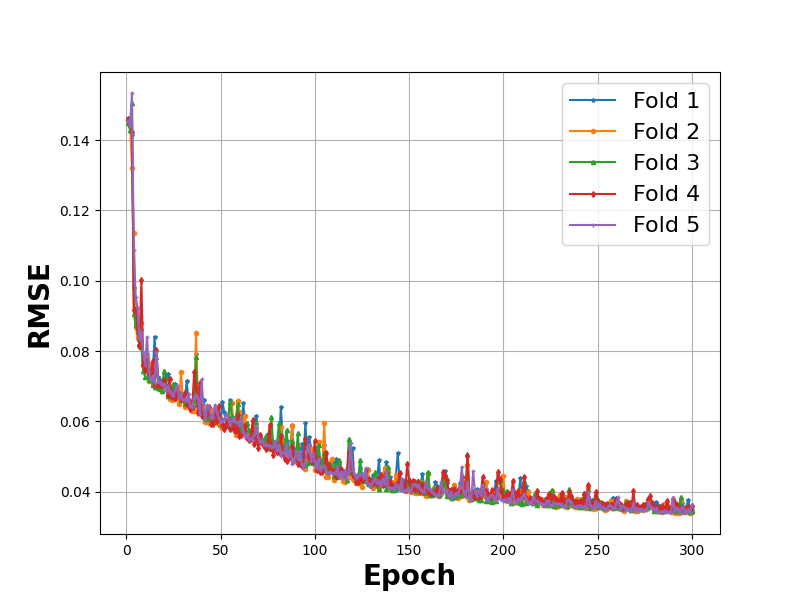}
        \caption{RRc Train RMSE K-folds}
        \label{fig:rrc_rmse_multi}
    \end{subfigure}
    \begin{subfigure}{0.475\textwidth}
        \includegraphics[width=\textwidth]{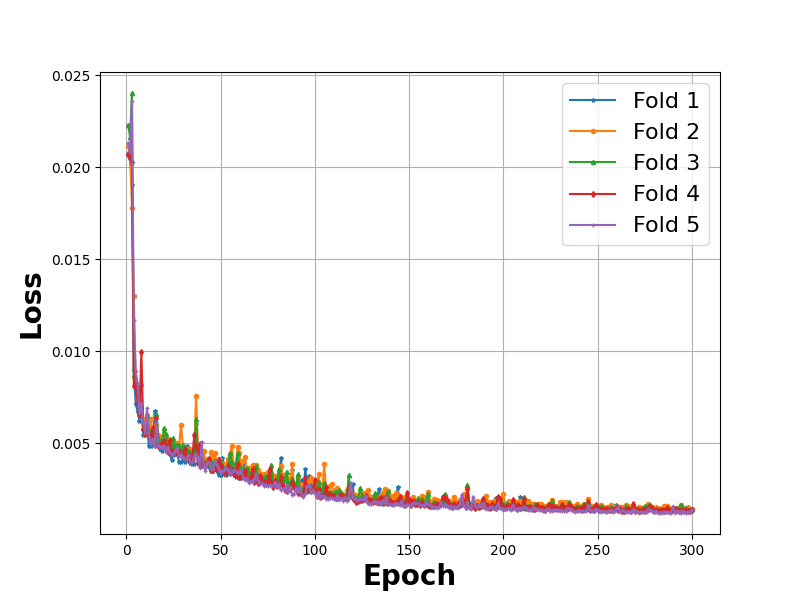}
        \caption{RRc Validation Loss K-folds}
        \label{fig:rrc_loss_multi_val}
    \end{subfigure}
    \centering
    \begin{subfigure}{0.475\textwidth}
        \includegraphics[width=\textwidth]{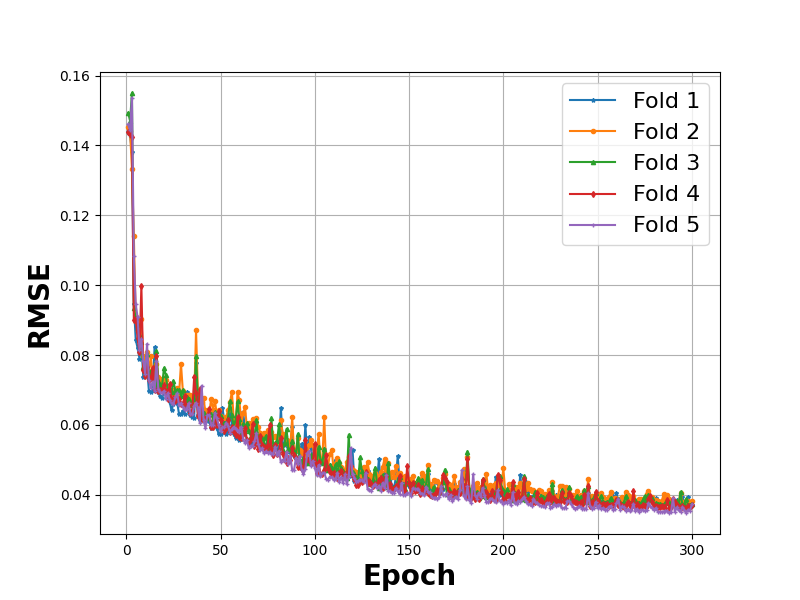}
        \caption{RRc Validation RMSE K-folds}
        \label{fig:rrc_rmse_multi_val}
    \end{subfigure}
  \end{minipage}
  \caption{\textbf{Training dynamics and convergence behavior of the RRc model.} Subfigures (a) and (b) illustrate the evolution of the loss and RMSE over training epochs, with separate curves for training and validation during a single run. Subfigures (c)-(f) show the corresponding metrics for each of the five folds during cross-validation.}
  \label{fig:rrc_convergence}
\end{figure}

\begin{figure}
  \begin{minipage}{1\textwidth}
    \centering
    \begin{subfigure}{0.475\textwidth}
        \includegraphics[width=\textwidth]{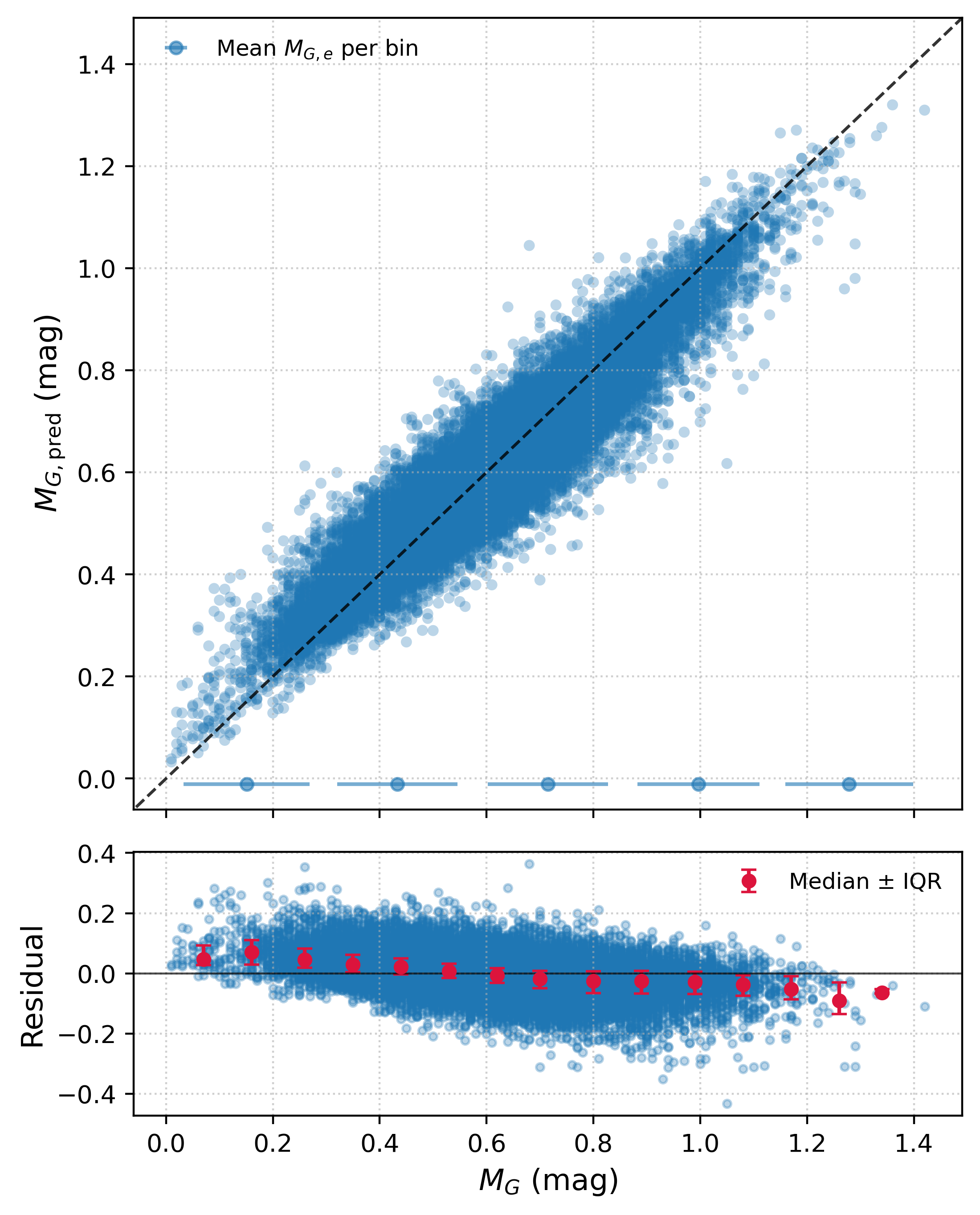}
        \caption{RRab Train}
        \label{fig:rrab_train}
    \end{subfigure}
    \begin{subfigure}{0.475\textwidth}
        \includegraphics[width=\textwidth]{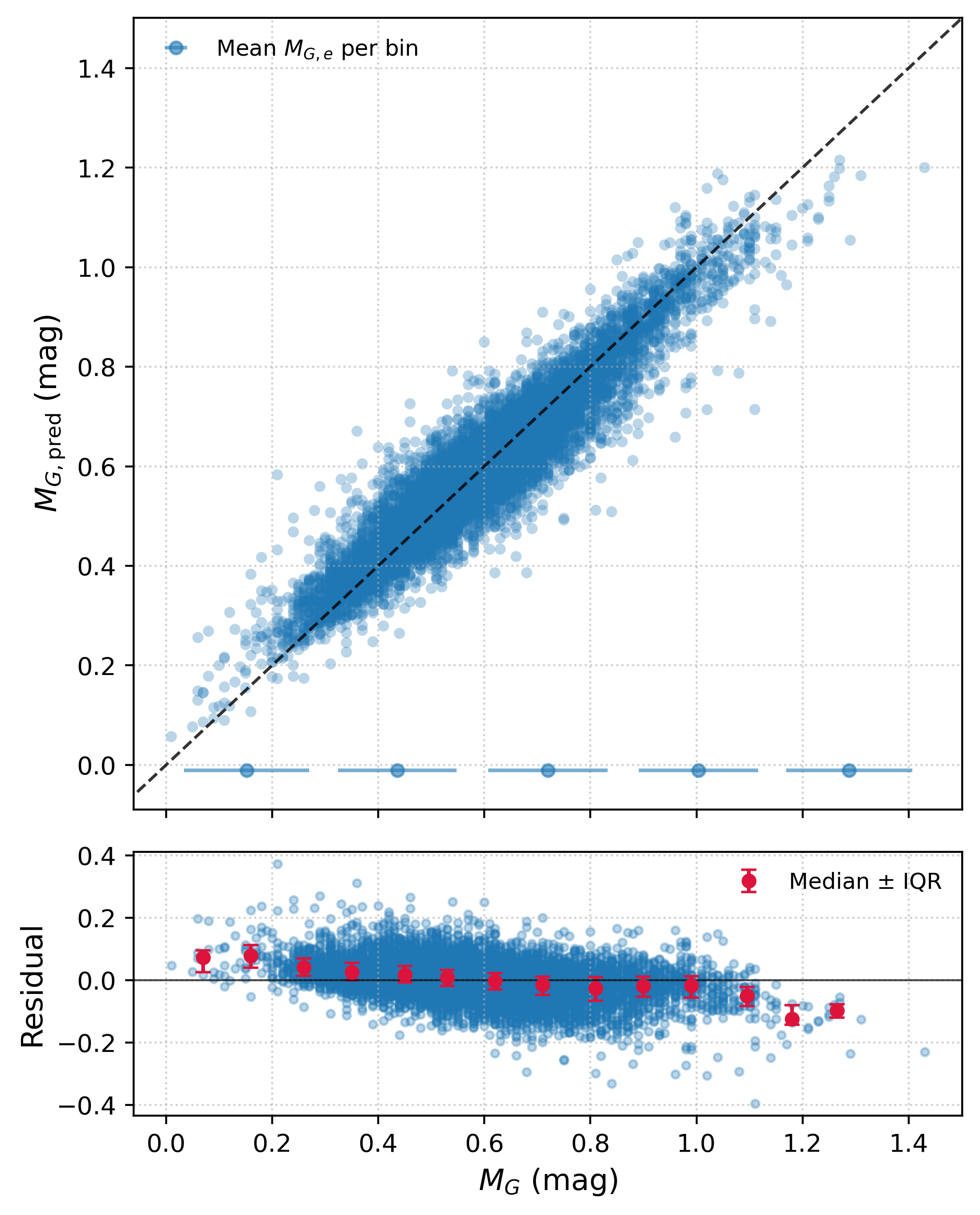}
        \caption{RRab Validation}
        \label{fig:rrab_valid}
    \end{subfigure}
    \begin{subfigure}{0.475\textwidth}
        \includegraphics[width=\textwidth]{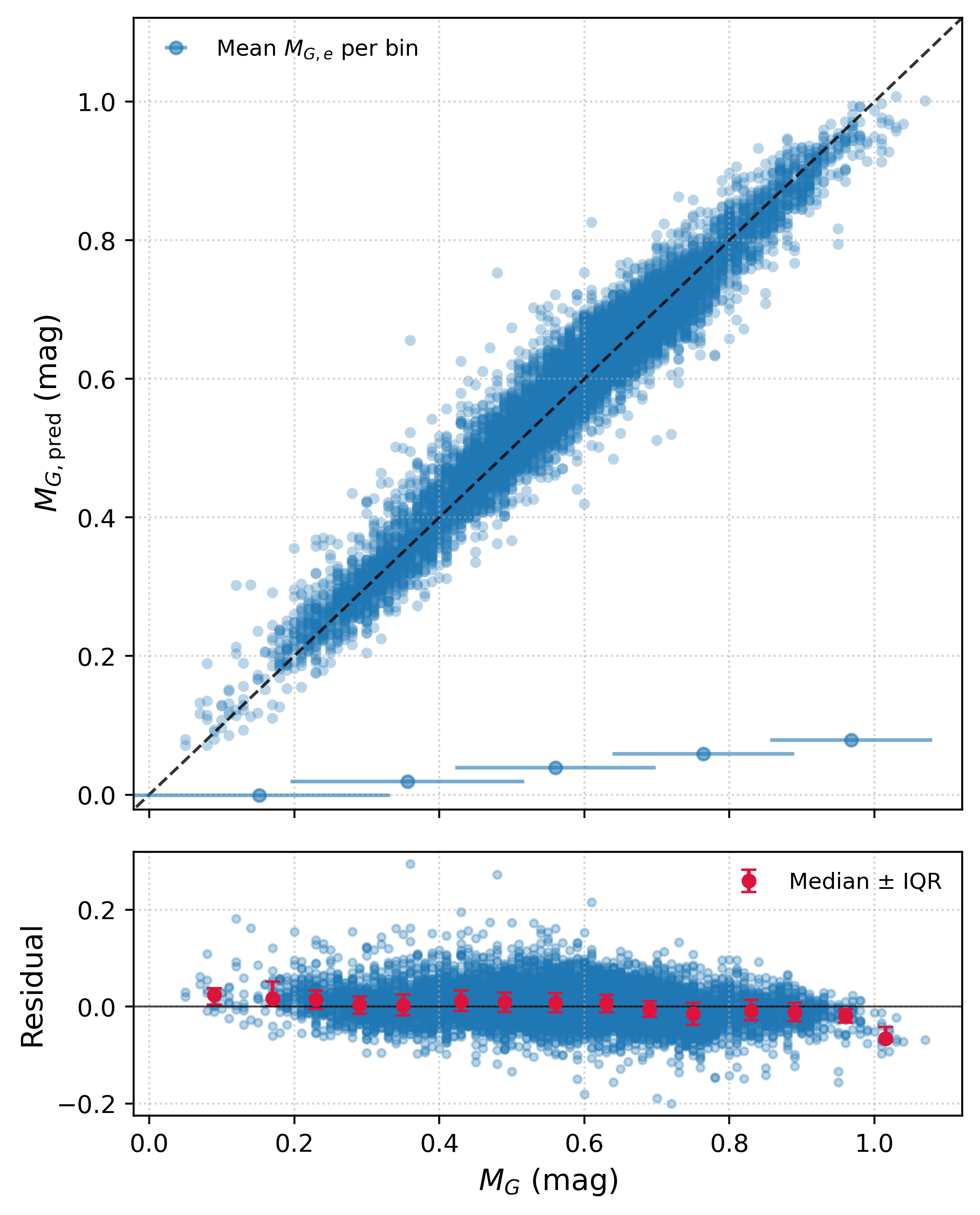}
        \caption{RRc Train}
        \label{fig:rrc_train}
    \end{subfigure}
    \centering
    \begin{subfigure}{0.475\textwidth}
        \includegraphics[width=\textwidth]{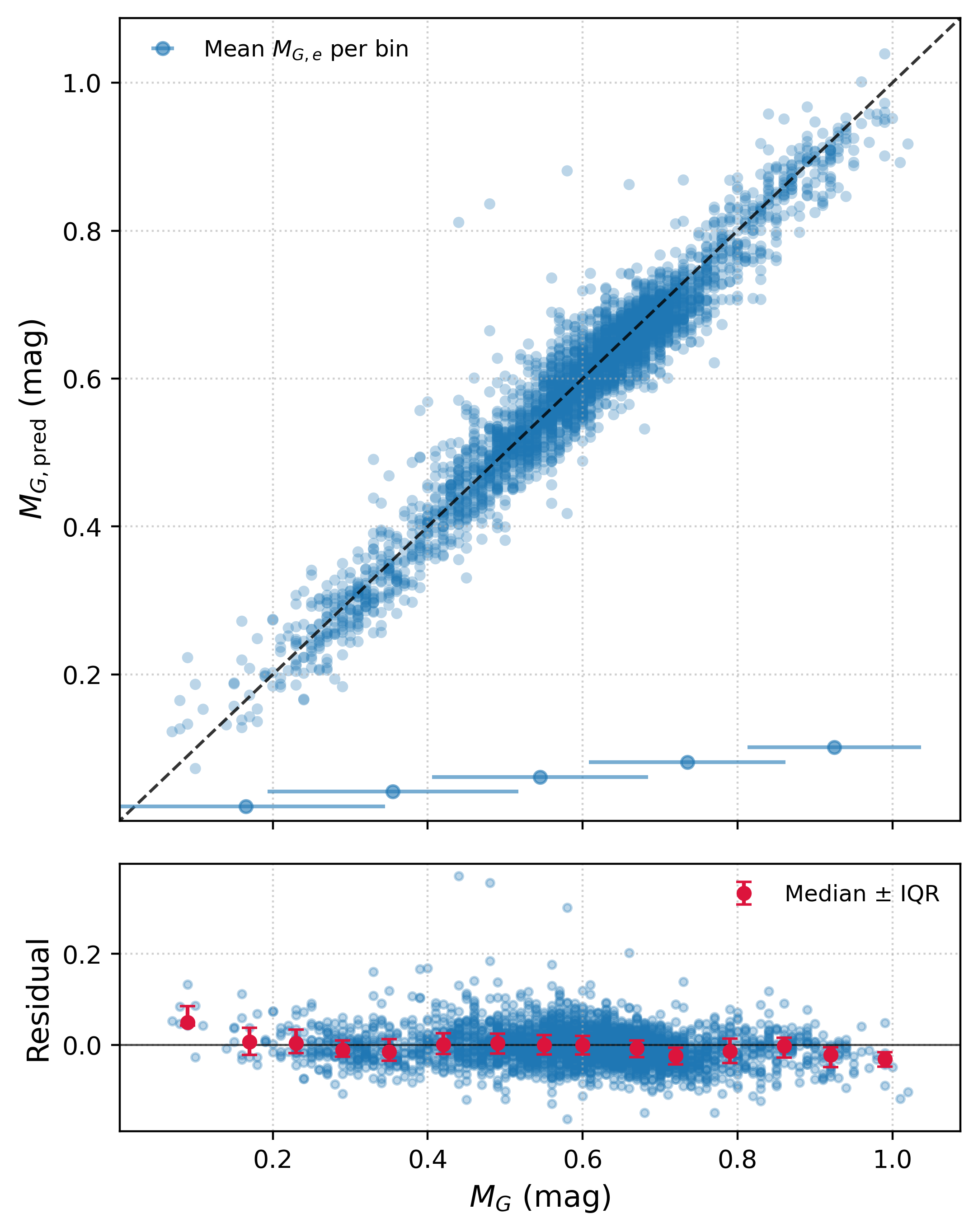}
        \caption{RRc Validation}
        \label{fig:rrc_valid}
    \end{subfigure}
  \end{minipage}
    \caption{\textbf{Regression results for RRab and RRc models.} Each subplot compares the label absolute magnitudes with the model predictions, with the black line representing the ideal one-to-one relation. Typical uncertainties of the training labels ($M_{G,\mathrm{e}}$) are shown in the upper subpanel, binned by $M_G$. Panels are arranged from top left to bottom right as RRab training, RRab validation, RRc training, and RRc validation sets. The lower subpanel in each plot shows residuals, where points indicate the median residual per $M_G$ bin and error bars represent the interquartile range (IQR).}
  \label{fig:ab_c_pred_mg_res}
\end{figure} 

\begin{figure*}[!htbp]
\begin{minipage}{1\textwidth}
\centering
\includegraphics[width=0.98\linewidth]{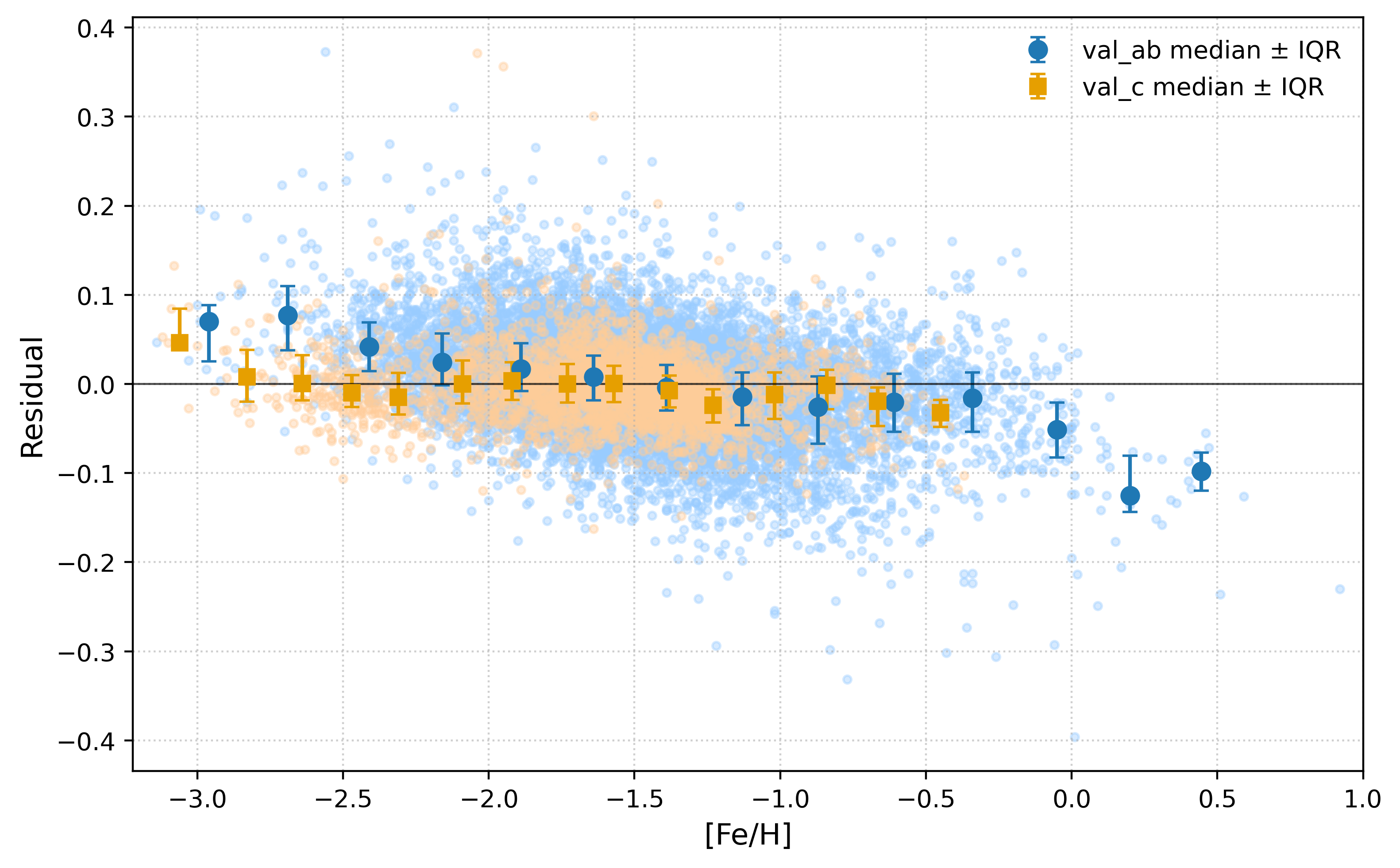}
\end{minipage}
\caption{\textbf{Validation residuals as a function of [Fe/H] for RRab and RRc models.} Residuals are defined as predicted minus label absolute magnitude ($M_{G\mathrm{,pred}} - M_G$). Blue and orange markers represent RRab and RRc stars, respectively. Points indicate the median residual within each [Fe/H] bin, with error bars denoting the inter-quartile range (IQR).}
\label{fig:residual_feh}
\end{figure*}

\begin{figure*}[!htbp]
\begin{minipage}{1\textwidth}
\centering
\includegraphics[width=0.98\linewidth]{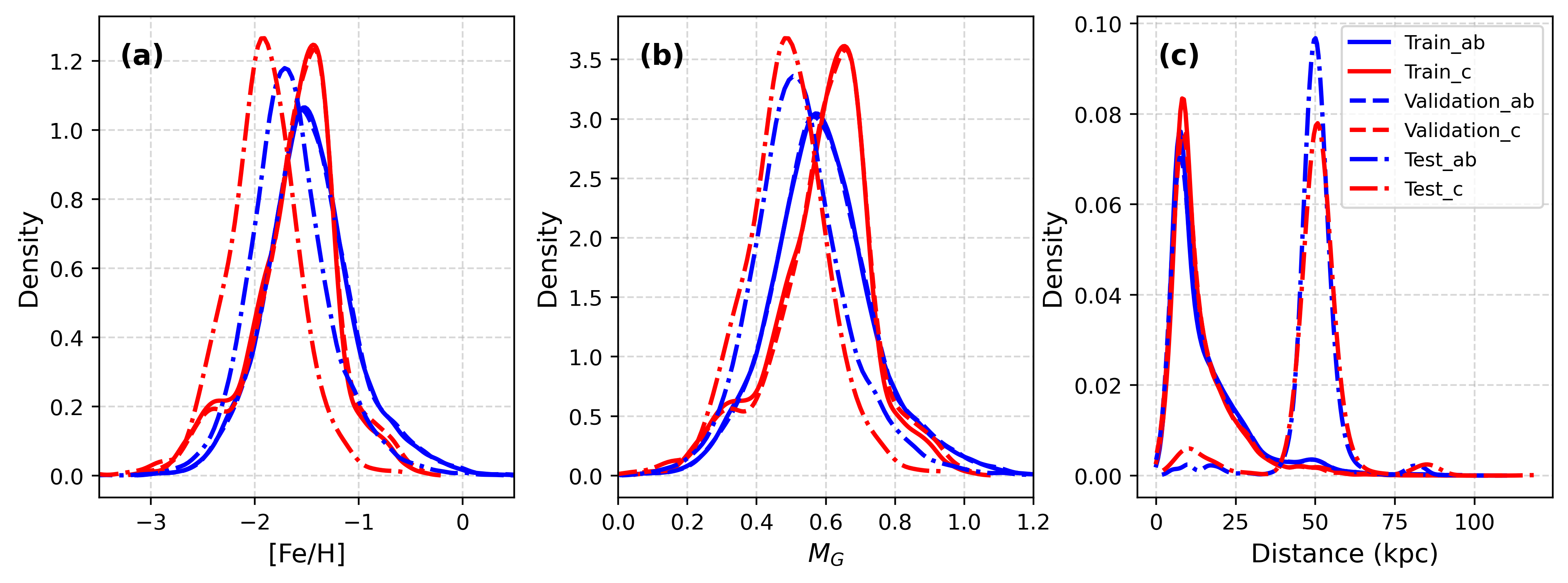}
\end{minipage}
\caption{\textbf{Distributions of key physical parameters for the training, validation, and test samples.} Panels (a), (b), and (c) show the kernel density estimates of [Fe/H], $M_G$, and distance, respectively. Solid, dashed, and dash-dotted lines represent the training, validation, and test sets. Blue and red correspond to RRab and RRc stars, respectively.}
\label{fig:para_distri}
\end{figure*}

\begin{figure*}[!htbp]
\begin{minipage}{1\textwidth}
\centering
\includegraphics[width=0.98\linewidth]{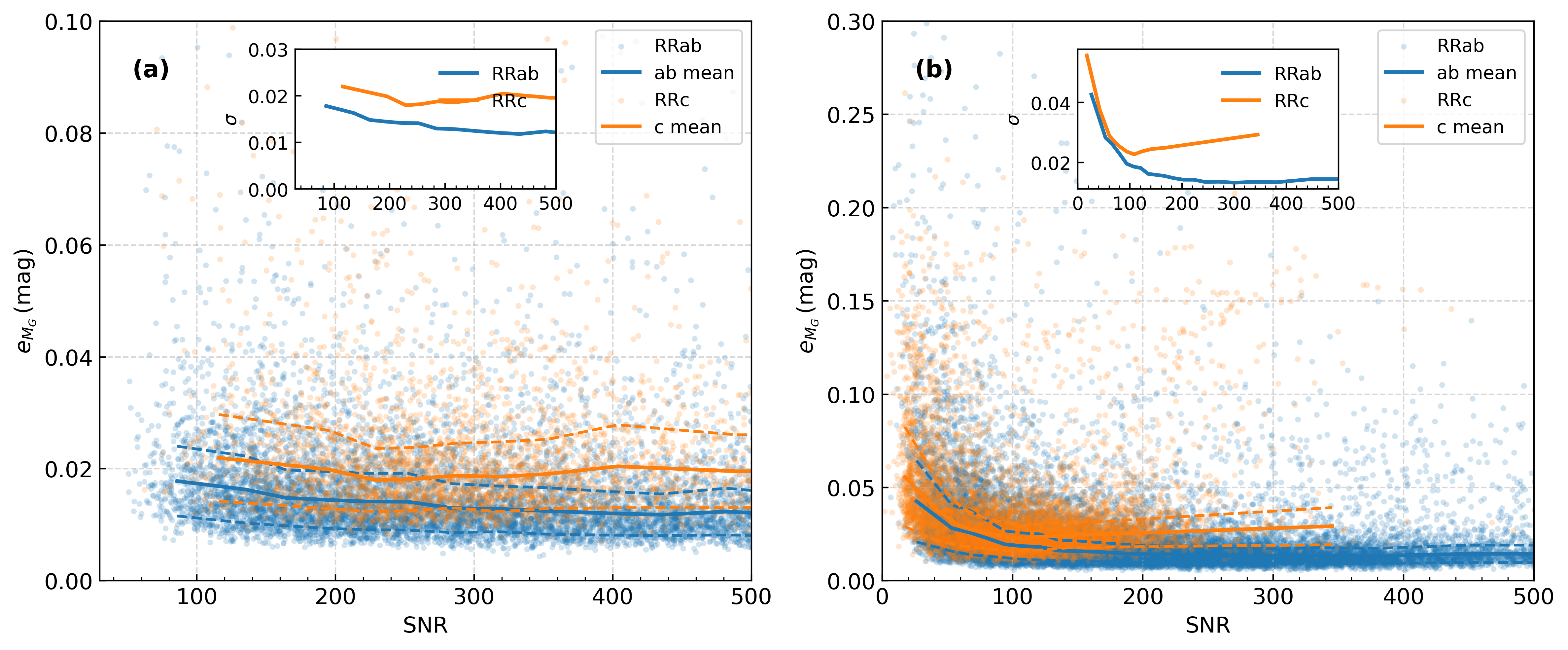}
\end{minipage}
\caption{\textbf{Dependence of prediction uncertainty on signal-to-noise ratio.} Panels (a) and (b) show the relation between SNR and $e_{M_G}$ for the validation set and the LMC sample, respectively, where $e_{M_G}$ is defined as the standard deviation of predicted magnitudes from 50 perturbed models. Solid and dashed lines indicate the mean and 1$\sigma$ dispersion within each SNR bin. The inset panels display the dispersion of $e_{M_G}$ as a function of SNR. Blue and orange correspond to RRab and RRc stars, respectively.}
\label{fig:SNR_M_G_e}
\end{figure*}

\newpage

\vspace*{30pt}

\noindent\textbf{SUPPLEMENTAL TABLES}
\vspace{2em}

\begin{table*}[!htbp]
\centering
\caption{Hyperparameter Details}
\begin{tabular}{c ccc ccc }
\hline
Model &  Dim1 & Dim2 & Loss Function & \# Epoch & Learning Rate \\
\hline 
RRab & 64 & 64 &  MSE & 300 & 0.0005 \\
RRc & 64 & 64 & Huber & 300 & 0.001 \\
\hline 
\end{tabular}
\label{tab:hyper}
\end{table*}

\begin{table}[!htbp]
\centering
\caption{Comparison of model performance across RRab and RRc subtypes}
\setlength{\tabcolsep}{15pt}
\begin{tabular}{lcccc}
\hline
Model & \multicolumn{2}{c}{RRab RMSE (mag) } & \multicolumn{2}{c}{RRc RMSE (mag) } \\
\cline{2-5}
& Train & Validation & Train & Validation \\
\hline
BiLSTM      & 0.0529 & 0.0534 & 0.0342 & 0.0364 \\
Linear Regression   & 0.1283 & 0.1293 & 0.0991 & 0.0971 \\
Random Forest & 0.0519 & 0.1407 & 0.0424 & 0.1093 \\
CNN         & 0.0703 & 0.0722 & 0.0501 & 0.0509 \\
RNN         & 0.0672 & 0.0681 & 0.0643 & 0.0623 \\
Transformer & 0.0476 & 0.0536 & 0.0338 & 0.0379 \\
\hline
\end{tabular}
\label{tab:model_comparison}
\end{table}

\vspace{20pt}

\newpage

% ============================================================
% SUPPLEMENTAL REFERENCES
% ============================================================
\noindent\textbf{SUPPLEMENTAL REFERENCES}
\vspace{-4em}  

\renewcommand{\refname}{}
\bibliographystyle{elsarticle-num}
\bibliography{innovation}